\documentclass[a4paper,10pt]{article}
\usepackage[utf8x]{inputenc}
\usepackage{graphicx}
\usepackage{amsmath,amssymb}
\usepackage{textcomp}
\usepackage[margin=3.0cm]{geometry}
\title{Ultrasensitivity and Fluctuations in the Barkai-Leibler Model of Chemotaxis Receptors in {\it Escherichia coli}}
\author{Ushasi Roy\textsuperscript{*}, Manoj Gopalakrishnan\textsuperscript{\textdied}
\\ \\
 Department of Physics, Indian Institute of Technology Madras,
 \\
 Chennai, Tamil Nadu, India
\\ \\
* ushasi@physics.iitm.ac.in \\
\textdied \; manojgopal@iitm.ac.in
}

\date{}
\begin{document}

\maketitle

\begin{abstract}
A stochastic version of the Barkai-Leibler model of chemotaxis receptors in \textit{Escherichia coli} is studied here 
with the goal of elucidating the effects of intrinsic network noise in their conformational dynamics. The model was originally 
proposed to explain the robust and near-perfect adaptation of \textit{E. coli} observed across a wide range of spatially uniform 
attractant/repellent (ligand) concentrations. In the model, a receptor is either active or inactive and can stochastically switch between the 
two states. The enzyme CheR methylates inactive receptors while CheB demethylates active receptors and the probability for a receptor 
to be active depends on its level of methylation and ligand occupation. In a simple version of the model with two methylation sites 
per receptor ($M=2$), we show rigorously, under a quasi-steady state approximation, that the mean active fraction of receptors is 
an ultrasensitive function of [CheR]/[CheB] in the limit of saturating receptor concentration. Hence the model shows zero-order 
ultrasensitivity (ZOU), similar to the classical two-state model of covalent modification studied by Goldbeter and Koshland (GK). 
We also find that in the limits of extremely small and extremely large ligand concentrations, the system reduces to two different 
two-state GK modules. A quantitative measure of the spontaneous fluctuations in activity is provided by the variance $\sigma_a^2$ 
in the active fraction, which  is estimated mathematically under linear noise approximation (LNA). It is found that $\sigma_a^2$ peaks 
near the ZOU transition. The variance is a non-monotonic, but weak function of ligand concentration and a 
decreasing function of receptor concentration. Gillespie simulations are also performed in models with $M=2,3$ and 4. 
For $M=2$, simulations show excellent agreement with analytical results obtained under LNA. Numerical results for $M=3$ and 
$M=4$ are qualitatively similar to our mathematical results in $M=2$; while all the models show ZOU in mean activity, the variance 
is found to be smaller for larger $M$. The magnitude of receptor noise deduced from available experimental data is consistent with our 
predictions. A simple analysis of the downstream signaling pathway shows that this noise is large enough to affect the motility 
of the organism, and may have a beneficial effect on it. The response of mean receptor activity to small time-dependent changes in 
the external ligand concentration is computed within linear response theory, and found to have a bilobe form, in agreement with 
earlier experimental observations. 

\end{abstract}

\section{Introduction}

The pioneering work \cite{bibGK} of Goldbeter and Koshland (GK) brought into light an interesting molecular switch-like transition, 
now referred to as Zero Order Ultrasensitivity (ZOU) \cite{bibZOU}, observed in reversible covalent modification (e.g. phosphorylation
or methylation) of a protein (substrate), catalyzed by two antagonistic enzymes. This switch-like behavior emerges in the limit where 
the substrate concentration is exceedingly large compared to the enzyme concentrations as well as their individual Michaelis constants.
In this ``zero-order" regime, the modified fraction of substrate, in a saturating environment, exhibits a sharp transition at a critical 
value of the ratio of the total concentrations of the antagonistic enzymes. Various aspects of GK switch has been studied over the years
\cite{bibBerg,  bibElf, bibXu}. The fluctuations associated with the ultrasensitive module has also been studied 
\cite{bibBerg, bibElf}. ZOU was also identified as allosteric cooperativity by Quian (2003) \cite{bibQian} and Ge and Quian (2008) \cite{bibGe}. 

The bacterium {\textit{E. coli} has thousands of receptors on its cell surface for a precise sensation of the extracellular environment. 
The two main types of chemotaxis receptors are Tar and Tsr. The receptor protein and the protein kinase CheA are linked 
by the linker protein CheW. The receptor protein, CheA and CheW function as a single signaling complex, which exists in \textit{active} or 
\textit{inactive} state, and can undergo stochastic switching between the states. The switching is regulated by methylation and ligandation 
(binding of chemoattractant/repellent) of the receptors. Methylation is carried out by the protein CheR while CheB demethylates receptors. 
In its active state, the receptor-CheW-CheA complex phosphorylates CheB and the response regulator CheY; while the former demethylates the 
receptor, the latter induces tumbles by binding to the flagellar motor. In this way, the internal biochemistry (methylation/demethylation) 
and external stimulus (attractant/repellent) regulates the swimming pattern of the organism (reviewed in \cite{bibMaini}). 

A remarkable property of chemotaxis in \textit{E. coli} is perfect adaptation to some chemoattractants. After exposure to a stimulus 
corresponding to an abrupt rise or fall in ligand concentration in a homogeneous environment, the frequency of tumbles returns to its 
pre-stimulus level after a time interval $\sim$ 4 seconds \cite{bibSegall}. A model of the methylation-demethylation cascade proposed by 
Barkai and Leibler (BL) \cite{bibBL} successfully explains this property using a few simple assumptions. The model was modified and extended 
in later years by other authors \cite{bibRao, bibPontius, bibMello, bibKollmann}. The important assumptions in the model are 
(a) CheR methylates inactive 
receptors while CheB (or its phosphorylated form, CheBp) demethylates active receptors, (b) the state of activity of a receptor depends 
on its methylation level, (c) binding of an attractant (ligand) molecule adversely affects the activity of a receptor, (d) ligand binding 
and dissociation are very fast compared to methylation and demethylation. Mello and Tu \cite{bibMello} showed that perfect adaptation is achieved 
only if the lowest methylation level is assumed to be always inactive, while the highest level is assumed to be always active. As a consequence 
of these assumptions, the mean receptor activity in the model in steady state turns out to be independent of the attractant concentration.  
Further, mathematical modeling has shown that, with suitable extensions, the model also successfully predicts the 
response of the network to a short-lived spike in attractant concentration\cite{bibReneaux}. 

In {\it E.coli}, the most abundant receptor, Tar, responding to the attractant methyl aspartate, has four methylation sites per monomer. 
For modeling purposes, it is convenient to represent  the methylation state of a single receptor by an index $m$, where $0\leq m\leq M$ 
with $M$ being the maximum value ($M=4$ for Tar). It is generally assumed that activation/inactivation as well as ligand binding/unbinding 
of a receptor happens much faster compared to methylation/demethylation. For $M=1$, the BL model is entirely identical with the Goldbeter-Koshland (GK) 
two-state system; the ligand concentration $L$ plays no role whatsoever in 
the dynamics. For $M\geq 2$, the intermediate state(s) which stochastically switches(switch) between active and inactive states modifies 
the dynamics; however, the model retains the same structure as the GK module (see Fig \ref{fig1}, where we have provided a schematic 
depiction of the BL model, for $M=2$). Therefore, it is pertinent to enquire whether ZOU is present also in the BL model. In the GK system 
($M=2$ in our model), fluctuations in activity are known to be large near the ZOU transition. In {\it E.coli} chemotaxis, intrinsic fluctuations 
are expected to play an important role. The flagellar motor responds ultrasensitively to [CheYp] (p denoting the phosphorylated form); 
the sensitive part of the motor response curve is less than 1$\mu$M in width \cite{bibCluzel,bibYuan_JMB}, hence it is likely that spontaneous and random fluctuations in activity are crucial in generating the run-and-tumble motion of the organism \cite{bibPontius}. For this reason, we believe 
that a detailed study of ZOU in the BL model, including a systematic treatment of fluctuations will make a useful addition to existing literature 
on ZOU as well as chemotaxis in {\it E.coli}. This is the motivation behind this paper. 

\begin{figure}[!h]
\begin{center}
 \includegraphics[width=10cm, angle = 0]{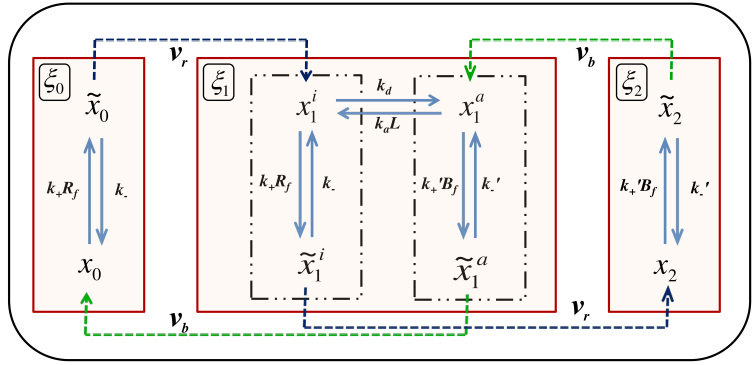}
\caption{A schematic figure of the 3-state BL Model showing all the biochemical reactions. The red boxes denote the three states;
$\xi_0$, $\xi_1$ and $\xi_2$ being the fractional concentrations of receptors in each. Blue dotted arrows with rates $\nu_r$ 
show the methylation process resulting in the increment in the state index while the green ones with rates $\nu_b$ depict demethylation with 
the decrement in the same.}
\label{fig1}
\end{center}
\end{figure}

\section{Materials and Methods}

\subsection{Model details}

Let us denote by $V$ the volume of a cell having a total of $N=A_{0}V$ receptor proteins (substrate), and two enzymes. In {\it E.coli}, 
these correspond to the methyltransferase CheR and methylesterase CheB, 
the total concentrations of each are denoted by $R_0$ and $B_0$ respectively. In the presently accepted model of signal transduction 
in {\it E.coli}, CheR methylates the receptor, while CheB demethylates it. 
According to the two-state model \cite{bibAsakura} the receptor exists 
in a complex form involving two more proteins, CheW and CheA, and the complex may exist in one of two 
conformational states, which we refer to as {\it active} and {\it inactive}. In the active state, CheA undergoes 
autophosphorylation and also phosphorylates the cytoplasmic messenger protein CheY. CheA phosphorylates CheB also, a process crucial in 
adaptation, but ignored here for the sake of maintaining close structural similarity with the GK system. 
The probability to be in the active state is an increasing function of the methylation level of the receptor. 
Chemoattractant molecules, if  present, also bind to the receptor, and this process increases the probability to find the receptor in 
inactive state (with chemorepellents, it is the reverse)\cite{bibMaini}. 

BL proposed a quantitative model \cite{bibBL} of the signal transduction cascade, based on the two-state model of Asakura and 
Honda \cite{bibAsakura}. In this model, CheR binds only to inactive receptors, while CheB binds only to active receptors \cite{bibEmonet}. 
In general, we denote by $M$ the total number of methylation sites. Let $A_m$ symbolically represent a receptor in methylation state $m$, 
and $\tilde{A}_m$ denotes its enzyme-bound form; the same symbols will also be used to denote the corresponding numbers (out of a total of $N$). 
The inactive and active versions are identified by superscripts $i$ and $a$ respectively. The binding rate of CheR to a receptor is $k_{+}$ and 
$k_{-}$ is the dissociation rate; for CheB, the corresponding quantities are $k_{+}^{\prime}$ and $k_{-}^{\prime}$ respectively. 
Attractant binding is relevant only for the intermediate state; we assume that the binding rate is $k_a$, dissociation rate is $k_d$ 
and attractant concentration is $L$ (assumed uniform). $\nu_r$ and $\nu_b$ represent the product formation rates from CheR and CheB-bound 
intermediate states, respectively. The concentrations of the free forms of CheR and CheB are denoted by $R_f$ and $B_f$ respectively. 
We also define a set of dissociation constants: $K_r=k_{-}/k_{+}, K_b=k_{-}^{\prime}/k_{+}^{\prime}, K_L=k_d/k_a$. 

If $\xi_m(t)$ denote the fraction of receptors in methylation state $m$ at time $t$, the total active fraction of receptors is given by 
$\xi_a(t)=\sum_{m=0}^{M}a_m(L)\xi_m(t)$, where $a_m(L)$ is the probability for a receptor in state $m$ to be active, $L$ being the (uniform) 
attractant concentration. We assume that the lowest state ($m=0$) is always inactive, while $m=M$ is always active; the intermediate state(s) 
can be either active or inactive, depending on whether it is attractant-bound. A simple mathematical form satisfying these conditions is 
$a_m(L)=a_m(0)K_m/(L+K_m)$ where $K_m$ is the dissociation constant for attractant binding to a receptor in state $m$
\cite{bibSourjik, bibKollmann, bibReneaux}. To satisfy the 
earlier assumption, we choose $K_{0}=0$ and $K_{M}=\infty$ (these are ideal values). We further choose $a_m^0=m/M$\cite{bibEmonet}, a simple 
mathematical form which satisfies our earlier assumptions.

\subsection{Fokker-Planck equation for $M=2$}

We will first carry out a detailed mathematical analysis of the properties of the BL model with $M=2$. Fig \ref{fig1} shows a 
schematic depiction of 
this case, showing all the biochemical reactions. In the limit where the modification rates $\nu_r$ and $\nu_b$ are very small 
in comparison with the other rates, it may be visualized as a combination of three weakly-coupled modules (the boxes in Fig~\ref{fig1}); 
the fractional populations of receptors in each module are $\xi_m$ with $m=0,1,2$; hence $\sum_{m=0}^{2}\xi_m=1$. In general, for a complete stochastic 
description of the dynamics of this system, we also need to keep track of time evolution of the intra-modular populations 
viz., $x_m$ and ${\tilde x}_m$. For $m=1$, there is a further sub-division: 
$x_1=x_1^{a}+x_1^{i}$ and ${\tilde x}_1={\tilde x}_1^{a}+{\tilde x}_1^{i}$, as explained in the previous section. For each $m$, 
$\xi_m=x_m+{\tilde x}_m$. It was shown by us in an earlier paper \cite{bibZOU} that these intra-modular fractions can be `integrated out' 
using one of the quasi-steady state approximation schemes (QSSA) \cite{bibSchnoerr}, whereby the replacement 
$x_i\to \overline{x_i}(\boldsymbol{\xi})$ naturally appears in the relevant inter-modular conversion rates. Here, we have introduced 
the compact vector symbol 
$\boldsymbol\xi\equiv (\xi_0,\xi_2)$ (because of the normalization constraint mentioned above, only two of the 
fractions $\xi_m$ are independent, we shall take these to be $\xi_0$ and $\xi_2$ purely for reasons of convenience). 

Therefore, within this approximation, the inter-modular dynamics may be expressed in terms of 
a probability density $P(\boldsymbol\xi;t)$, which satisfies a non-linear Fokker-Planck equation (FPE) in the form of a local equation of continuity\cite{bibZOU}:

\begin{equation} \label{FPE}
 \frac{\partial P(\boldsymbol\xi,t)}{\partial t} = -\frac{\partial}{\partial \boldsymbol\xi}\cdot \bf{J}
 \end{equation}

where $\textbf{J} \equiv \{J_0, J_2\}$ with $J_m = v_mP-\frac{\partial}{\partial \xi_m}(D_m P)$ (for $m=0,2$) denoting the components of the state space probability current ``vector''. Here, the drift and diffusion coefficients are given by 

\begin{equation} \label{v}
v_m= \delta (\omega_{1m} - \omega_{m1})~~;~~~D_m=  \frac{\delta^2}{2} (\omega_{1m} + \omega_{m1})~~;~~~m=0,2
\end{equation}

with $\delta=N^{-1}$, $\omega_{10} = N \nu_b \tilde{x}_1^a $,  $\omega_{01} = N \nu_r \tilde{x}_0$, $\omega_{12} = N \nu_r \tilde{x}_1^i $ and $\omega_{21} = N \nu_b \tilde{x}_2 $. 

Incorporating the rates $\omega_{mn}$ in the above expressions for $v_m$ and $D_m$ and employing the inter-module dynamics under the 
standard quasi-steady state approximation (sQSSA) as discussed in \ref{S1_Appendix}, we arrive at the following expressions for the drift and diffusion 
coefficients: 

\begin{eqnarray}\label{V}
 v_0(\boldsymbol{\xi}) &=&  \frac{\nu_b B_f}{B_f+K_b}\left(\frac{K_L}{L+K_L}\right)(1-\xi_0-\xi_2) - \frac{\nu_r R_f}{R_f+K_r}\xi_0 , \nonumber \\
 v_2(\boldsymbol{\xi}) &=&  \frac{\nu_r R_f}{R_f+K_r}\left(\frac{L}{L+K_L}\right)(1-\xi_0-\xi_2) - \frac{\nu_b B_f}{B_f+K_b}\xi_2, 
\end{eqnarray}

and

\begin{eqnarray}\label{D}
 D_0(\boldsymbol{\xi}) &=& \frac{1}{2N} \Bigg[ \frac{\nu_b B_f}{B_f+K_b}\left(\frac{K_L}{L+K_L}\right)(1-\xi_0-\xi_2) + \frac{\nu_r R_f}{R_f+K_r}\xi_0 \Bigg], \nonumber \\
 D_2(\boldsymbol{\xi}) &=& \frac{1}{2N} \Bigg[ \frac{\nu_r R_f}{R_f+K_r}\left(\frac{L}{L+K_L}\right)(1-\xi_0-\xi_2) + \frac{\nu_b B_f}{B_f+K_b}\xi_2 \Bigg].
\end{eqnarray}
}

The free enzyme concentrations $R_f$ and $B_f$ are also functions of $\xi_0,\xi_2$, and the expressions are given by (see \ref{S1_Appendix})

\begin{eqnarray} \label{RB}
R_f(\boldsymbol{\xi}) &=& \frac{R_0K_r(L+K_L)}{K_r(L+K_L)+A_0[K_L\xi_0+L(1-\xi_2)]}  ~ , \nonumber \\
B_f(\boldsymbol{\xi}) &=& \frac{B_0K_b(L+K_L)}{K_b(L+K_L)+A_0[K_L(1-\xi_0)+L\xi_2]}  ~ .
\end{eqnarray}



\section{Results}

\subsection{Averages as fixed points}

From (\ref{FPE}), it is seen after the required averaging that the time evolution of the mean receptor fractions is given by 
the vector equation
\begin{equation}\label{NEW}
\frac{d \overline{ \bf{\xi} } } {dt} = \overline{\bf{v}(\boldsymbol{\xi})},
\end{equation}
where ${\bf v}(\boldsymbol{\xi})\equiv (v_0(\boldsymbol{\xi}),v_2(\boldsymbol{\xi}))$ is the drift ``vector''. In general, this vector 
vanishes at one or more points 
in the $(\xi_0,\xi_2)$ space; following van Kampen\cite{bibVan}, we shall refer to this point (assuming only one exists) as the fixed point, 
denoted  $\boldsymbol\xi^*=(\xi_0^*,\xi_2^*)$. It can be shown (see later) that in the large $N$-limit, the fixed point becomes identical to 
the mean value $\overline{{\bf\xi}}$. 

Under conditions $R_f\ll K_r$ and $B_f\ll K_b$, which we shall assume to hold for reasons of simplicity, it follows from (\ref{V}) that the fixed point is given by 

\begin{eqnarray} \label{fxd_xi}
 \xi_0^* = \frac{B_f^{*2} K_r^2 \nu_b^2 K_L}{(B_f^* K_r \nu_b + R_f^* K_b \nu_r)(B_f^* K_r \nu_b K_L + R_f^* K_b \nu_r L)} \nonumber \\
 \xi_2^* = \frac{R_f^{*2} K_b^2 \nu_r^2 L}  {(B_f^* K_r \nu_b + R_f^* K_b \nu_r)(B_f^* K_r \nu_b K_L + R_f^* K_b \nu_r L)}, 
\end{eqnarray}

where $R_f^*\equiv R_f(\boldsymbol{\xi}^*)$ and $B_f^*\equiv B_f(\boldsymbol{\xi}^*)$ are given by (\ref{RB}), with the replacement 
$\boldsymbol{\xi}\to\boldsymbol{\xi}^*$ in the right hand side of the equations. (\ref{RB}) and (\ref{fxd_xi}) implicitly give the fixed point $\boldsymbol{\xi}^*$. 

\subsection{Evaluation of the covariance matrix from the linear FPE}

\noindent Next, we expand $v_m$ and $D_m$ in Taylor Series about the fixed point values. Let us define the deviation
 $\boldsymbol\xi^{\prime}=\boldsymbol\xi-\boldsymbol\xi^*$; we also define its probability distribution $\Phi(\boldsymbol\xi^{\prime})\equiv 
 P(\boldsymbol\xi^*+\boldsymbol\xi^{\prime})$, which satisfies the equation

\begin{equation}\label{FPE1}
\frac{\partial \Phi(\boldsymbol\xi^{\prime},t)}{\partial t} = -\frac{\partial}{\partial \boldsymbol\xi^{\prime}}\cdot \bf{J}^{\prime}
\end{equation}

where $\bf{J}^{\prime}(\boldsymbol\xi^{\prime})\equiv \bf{J}(\boldsymbol\xi^*+\boldsymbol\xi^{\prime})$, and may be expanded as follows: 

\begin{equation}
 J^{\prime}_m(\boldsymbol\xi^{\prime}) = \Phi(\boldsymbol\xi^{\prime})\Big[v_m(\boldsymbol\xi^*)+\sum_n\xi'_n\frac{\partial v_m}{\partial \xi_n}
 \Big\rvert_{\boldsymbol\xi^*}+...\Big]-
 \frac{\partial}{\partial \xi_m^{\prime}}\Big[\Phi(\boldsymbol\xi^{\prime})\Big(D_m(\boldsymbol\xi^*)+\sum_l\xi'_l\frac{\partial D_m}{\partial \xi_l}
 \Big\rvert_{\boldsymbol\xi^*}+...\Big)\Big]
\end{equation}

\noindent Keeping the leading term in each, and noting that ${\bf v}(\boldsymbol{\xi}^*)=0$, we arrive at (for $m$=0,2)

\begin{equation}\label{current}
J^{\prime}_m \simeq \Phi(\boldsymbol\xi^{\prime})\sum_{n=0,2} \beta_{mn}\xi^{\prime}_n - D_m^{*}\frac{\partial \Phi(\boldsymbol\xi^{\prime})}{\partial \xi^{\prime}_m}
\end{equation}

where $\beta_{mn} = \frac{\partial v_{m}} {\partial \xi_n}\Big|_{\boldsymbol\xi^*}$, $D_m^*\equiv D_m(\boldsymbol\xi^*)$. 
Substitution of (\ref{current}) in (\ref{FPE1}) leads to the multivariate linear Fokker-Planck 
equation (LFPE) \cite{bibVan}

\begin{equation} \label{LinFPE}
 \frac{\partial \Phi (\boldsymbol\xi^{\prime},t)}{\partial t} = -\sum_{m=0,2}\frac{\partial}{\partial \xi_m^{\prime}}\Big[ \Phi\sum_{n=0,2} 
 \beta_{mn}\xi^{\prime}_n\Big] + \sum_{m=0,2} D_m^{*} \frac{\partial^2 \Phi}{\partial {\xi_m^{\prime}}^2},
\end{equation}

Let us now define the covariances $\sigma_{mn}=\langle \xi^{\prime}_m\xi^{\prime}_n\rangle$ for $m,n=0,2$ (for $m=n$ 
these become the variances), which are evaluated in a convenient way by defining the moment generating function 
(Fourier transform) 

\begin{equation}
G(\boldsymbol\mu,t)=\int_{-\infty}^{\infty}d\xi^{\prime}_0d\xi^{\prime}_2 e^{-i\boldsymbol\mu\cdot\boldsymbol\xi^{\prime}}\Phi(\boldsymbol\xi^{\prime},t)
\end{equation}

where $\boldsymbol\mu=(\mu_0,\mu_2)$.The moment generating function admits the Taylor expansion

\begin{equation} \label{mGen}
G(\boldsymbol\mu,t)=1 - i\boldsymbol\mu\cdot\overline{\boldsymbol\xi^{\prime}(t)}-\frac{1}{2}\sum_{m,n=0,2}\mu_m\mu_n\sigma_{mn}(t)+.........
\end{equation}

Since $\Phi(\boldsymbol\xi^{\prime})$ satisfies the LFPE given by (\ref{LinFPE}), the moment generating function satisfies the equation

\begin{equation} \label{lFPE}
\frac{\partial G(\boldsymbol{\mu},t)}{\partial t}=\sum_{m,n=0,2}\beta_{mn}\mu_m\frac{\partial G}{\partial \mu_n}-G\sum_{m=0,2}D_m^*\mu_m^2
\end{equation}

Under stationary conditions ($t\to\infty$), the left hand side of (\ref{lFPE}) becomes zero; next we substitute (\ref{mGen}) in (\ref{lFPE}), 
and find that $\overline{\boldsymbol\xi^{\prime}}=0$ in the long-time limit. Therefore, within LNA, the fixed point values are equal to the 
statistical means of the respective quantities. The steady state covariance matrix satisfies the Lyapunov equation\cite{bibVan}

\begin{equation}\label{Lyapunov}
\boldsymbol\beta \boldsymbol\sigma + \boldsymbol\sigma \boldsymbol\beta^{T} + 2\boldsymbol D=0
\end{equation}

where $D_{mn}=D^*_m\delta_{mn}$ are the elements of the (diagonal) diffusion matrix $\boldsymbol{D}$. The following expressions follow from (\ref{Lyapunov}):

\begin{eqnarray}
 \sigma_{00} &=&  -{\mathcal D}^{-1}\Big[ \Big(\beta_{22}(\beta_{00}+\beta_{22})-\beta_{02}\beta_{20}\Big)D_0^*+\beta_{02}^2D_2^* \Big] \nonumber \\
 \sigma_{22} &=& -{\mathcal D}^{-1}\Big[ \beta_{20}^2D_0^*+\Big(\beta_{00}(\beta_{00}+\beta_{22})-\beta_{02}\beta_{20}\Big)D_2^* \Big] \nonumber \\
 \sigma_{02} &=& ~~{\mathcal D}^{-1}\Big[ \beta_{20}\beta_{22}D_0^* + \beta_{00}\beta_{02}D_2^* \Big]
\end{eqnarray}
where ${\mathcal D}=  (\beta_{00}+\beta_{22})(\beta_{00}\beta_{22}-\beta_{02}\beta_{20})$. Explicit expressions for the coefficients $\beta_{mn}$ are to be found in \ref{S2_Appendix}. 

\subsection{Mean and fluctuations in Activity}

The active fraction of receptors in the present model is given by 

\begin{equation}\label{ACTIVE}
\xi_a = \xi_1^{a} + \xi_2 = \frac{(1-\xi_0)}{1+\ell}+\frac{\ell}{1+\ell} \xi_2.
\end{equation}

where $\ell=L/K_L$. Within LNA, the average fractions $\overline{\xi}_m=\xi_m^*$, hence $\overline{\xi}_a=\xi_a^*$, where $\xi_a^*$ is given by 
(\ref{ACTIVE}), with the replacements $\xi_m\to\xi_m^*$ in the right hand side and $\xi_m^*$ given by (\ref{fxd_xi}).

The variance of the active fraction is given by $\sigma_a^2=\overline{\xi_a^2}-\overline{\xi}_a^2$, which is also computed using 
(\ref{ACTIVE}). In terms of the covariances $\sigma_{mn}$, this is given by 

\begin{equation}\label{VARIANCE}
\sigma_a^2=\frac{\sigma_{00}+\ell^2\sigma_{22}-2\ell\sigma_{02}}{(1+\ell)^2}. 
\end{equation}

\subsection{Ultrasensitivity in the mean activity}

We will now explore the large $A_0$ limit of the fixed point. The fixed point values $\xi_0^*$ and $\xi_2^*$ given by (\ref{fxd_xi}) 
can be expressed alternatively as 

\begin{eqnarray} \label{xi0xi2}
\xi_0^* = \left[1 + \Big(\frac{R_f^*}{B_f^*}\Big)^2 \frac{K_b^2\nu_r^2\ell}{K_r^2\nu_b^2} + \Big(\frac{R_f^*}{B_f^*}\Big)\frac{K_b\nu_r(\ell+1)}
{K_r\nu_b}  \right]^{-1} \nonumber \\
\xi_2^* = \left[1 + \Big(\frac{B_f^*}{R_f^*}\Big)^2 \frac{K_r^2\nu_b^2}{K_b^2\nu_r^2\ell} + \Big(\frac{B_f^*}{R_f^*}\Big)\frac{K_r\nu_b(\ell+1)}
{K_b\nu_r\ell} \right]^{-1}
\end{eqnarray}

In order to understand the behavior of the above expressions in the $A_0\to\infty$ limit, let us conjecture expansions of the form

\begin{eqnarray} \label{xi_m}
\xi_m^*=\xi_m^{(0)}+\frac{1}{A_0}\xi_m^{(1)} + \frac{1}{A_0^2}\xi_m^{(2)} + . . .
\end{eqnarray}

Consider now the expressions for $R_f$ and $B_f$ as given in (\ref{RB}); after using the expansions given by (\ref{xi_m}), we find 
the following asymptotic forms  in the large $A_0$ limit:

\begin{eqnarray}
R_f & \sim & \frac{1}{A_0} R_f^{(1)} + \frac{1}{A_0^2} R_f^{(2)} + . . . \nonumber \\
B_f & \sim & \frac{1}{A_0} B_f^{(1)} + \frac{1}{A_0^2} B_f^{(2)} + . . .
\end{eqnarray}

where 

\begin{eqnarray} \label{Rf12Bf12}
R_f^{(1)} & = & \frac{R_0K_r(\ell+1)}{\xi_0^{(0)}+\ell(1-\xi_2^{(0)})}  \nonumber \\
R_f^{(2)} & = & - \frac{R_0K_r(\ell+1)[K_r(\ell+1)+\xi_0^{(1)}-\ell\xi_2^{(1)}]}{[\xi_0^{(0)}+\ell(1-\xi_2^{(0)})]^2} \nonumber \\
B_f^{(1)} & = & \frac{B_0K_b(\ell+1)}{1-\xi_0^{(0)}+\ell\xi_2^{(0)}}  \nonumber \\
B_f^{(2)} & = & - \frac{B_0K_b(\ell+1)[K_b(\ell+1)-\xi_0^{(1)}+\ell\xi_2^{(1)}]}{[1-\xi_0^{(0)}+\ell\xi_2^{(0)}]^2}.
\end{eqnarray}

The ratio $R_f/B_f$, upto ${\mathcal O}(1/A_0)$, is given by 

\begin{eqnarray} \label{Rf/Bf}
\frac{R_f}{B_f} & = & \frac{R_f^{(1)}}{B_f^{(1)}} \Big[ 1+\frac{1}{A_0} \Big( \frac{R_f^{(2)}}{R_f^{(1)}}-\frac{B_f^{(2)}}{B_f^{(1)}} 
\Big) + . . . \Big] 
\end{eqnarray}

Substituting (\ref{Rf/Bf}) in (\ref{xi0xi2}), we find 

\begin{eqnarray} \label{xi00}
\xi_0^{(0)} &=& \Bigg[ 1 + \frac{R_f^{(1)}}{B_f^{(1)}} \frac{K_b\nu_r}{K_r\nu_b} \Big[ \frac{R_f^{(1)}}{B_f^{(1)}}\frac{K_b\nu_r}
{K_r\nu_b}\ell + (\ell+1) \Big] \Bigg]^{-1} \nonumber \\
\xi_2^{(0)} &=& \Bigg[  1 + \frac{B_f^{(1)}}{R_f^{(1)}} \frac{K_r\nu_b}{K_b\nu_r} \Big[ \frac{B_f^{(1)}}{R_f^{(1)}}\frac{K_r\nu_b}
{K_b\nu_r}\frac{1}{\ell} + \frac{\ell+1}{\ell} \Big] \Bigg]^{-1}
\end{eqnarray}

which give the leading terms in the expansions in (\ref{xi_m}), and are valid for arbitrary $\ell$. In order to take the analysis further, 
we study the limits $\ell\to0$ and $\ell\to\infty$ separately. 

\subsubsection{Small $\ell$ expansion}

Consider now the limit of small ligand concentrations, $\ell\ll 1$. We assume that, in this case, the $zero^{th}$ order terms 
$\xi_m^{(0)}$ in (\ref{xi00}) can be expanded as 

\begin{eqnarray} \label{sl}
\xi_m^{(0)} & = & \xi_m^{00}+\ell\xi_m^{01} + \ell^2\xi_m^{02}+...
\end{eqnarray}

From (\ref{Rf12Bf12}), we have

\begin{equation} \label{xxo}
\frac{R_f^{(1)}}{B_f^{(1)}}=\frac{R_0K_r}{B_0K_b} \Bigg[ \frac{1-\xi_0^{(0)}+\ell\xi_2^{(0)}}{\xi_0^{(0)}+\ell(1-\xi_2^{(0)})} \Bigg]
\end{equation}

Substituting the small $\ell$ expansions of $\xi_m^{(0)}$ as given in (\ref{sl}), we obtain

\begin{equation} \label{R1B1}
\frac{R_f^{(1)}}{B_f^{(1)}}\Bigg\rvert_{\ell=0}=\frac{R_0K_r}{B_0K_b} \Bigg[ \frac{1-\xi_0^{00}}{\xi_0^{00}} \Bigg]
\end{equation}

Substituting (\ref{R1B1}) in (\ref{xi00}) leads to the equations

\begin{equation} \label{xi0020_sL}
 \xi_0^{00}(\xi_0^{00}-1)(1-\alpha)=0~~~;~~~\xi_2^{00}=0.
\end{equation}

where 

\begin{equation}\label{alpha}
\alpha = \frac{R_0\nu_r}{B_0\nu_b}
\end{equation} 

is the control parameter that characterizes ZOU. The implications are (a) the population in the highest methylation state is vanishingly small 
in the limits $\ell\to 0$ and $A_0\to\infty$ (b) $\xi_0^{00}=0$ or 1 unless $\alpha=1$: this points to a jump-like behavior for $\xi_0$ 
(and therefore, $\xi_1$, considering that $\xi_0^{00}+\xi_1^{00}=1$) in these limits. The 'critical point' $\alpha_c=1$ is the same as 
what was derived in the seminal paper of Goldbeter and Koshland\cite{bibGK}, and referred to as the GK point in our earlier paper \cite{bibZOU}. 

\subsubsection{Large $\ell$ expansion}

We next consider a large $\ell$ expansion of the $zero^{th}$ order term $\xi_m^{(0)}$ of large $A_0$ expansion
in the following form
\begin{eqnarray}
\xi_m^{(0)} & = &\underline{\xi}_m^{00}+\frac{1}{\ell}\underline{\xi}_m^{01} + \frac{1}{\ell^2}\underline{\xi}_m^{02}+...
\end{eqnarray}

Substituting these in (\ref{xxo}), we obtain

\begin{equation} \label{Rf/Bf_largeL}
\frac{R_f^{(1)}}{B_f^{(1)}}\Bigg\rvert_{\ell \to\infty}=\frac{R_0K_r}{B_0K_b} \Bigg[ \frac{\underline{\xi}_2^{00}}{1-\underline{\xi}_2^{00}} \Bigg]
\end{equation}

Therefore, after substitution of (\ref{Rf/Bf_largeL}) in (\ref{xi00}), we find 

\begin{equation} \label{xi0020_lL}
\underline{\xi}_0^{00}=0~~~;~~~\underline{\xi}_2^{00}\bigg(\underline{\xi}_2^{00}-1\bigg)(\alpha-1)=0
\end{equation}

which mirror (\ref{xi0020_sL}) derived in the opposite, $\ell\to 0$ limit. In the present case, it is $\xi_2$ that displays the jump 
transition from 1 to 0 as $\alpha$ crosses 1 (and $\xi_1$ does the reverse), while $\xi_0$ is vanishingly small at all $\alpha$. 
These analytical results, given in equations (\ref{xi0020_sL}) and (\ref{xi0020_lL}) are supported by numerical simulations, to be discussed in the following section.

\section{Stochastic Gillespie simulations}
For $M=2$, we simulated the reaction scheme given in Fig (\ref{fig1}) using Gillespie algorithm \cite{bibGillespie}. The specific numerical
values for various parameters correspond to the methylation–demethylation reactions of chemotaxis receptors in
the bacterium \textit{E. coli}, previously used by various authors \cite{bibZOU, bibReneaux}. Initially, we choose all the receptors to be in 
inactive, unbound configuration at the lowest methylation/inactive level ($m=0$). The system then evolves with the preassigned parameters 
(refer to Table \ref{table1})
and eventually reaches its steady state where the mean receptor number in each methylation level becomes time-independent. 
BL model with $M=2$ comprises of 8 conformational states and the total number of possible biochemical reactions 
is 9 (5 reversible and 4 irreversible reactions). 

Similarly for $M=3$, there are 12 
configuration states and 14 biochemical reactions while for $M=4$, these numbers become 
16 and 19 respectively. Once it is ensured that the system has reached its steady state, we study the average and variance of the active 
receptor fraction $\xi_a$ by varying $R_0$, $N$, $L$ and finally $M$, while $B_0$ was kept constant throughout. The fractions of receptors 
in different methylation levels were also kept track of. 

\begin{table}[!ht]
\centering
\setlength{\tabcolsep}{7 pt}
\renewcommand{\arraystretch}{1.5}
\caption{
{\bf A list of values of the various parameters used in numerical simulations} (previously used in \cite{bibZOU} and \cite{bibReneaux}).
The experimental values for the concentration of CheR in {\it E.coli} can be found in Table \ref{table2} in \ref{S3_Appendix} but in our simulations, we have varied it in the range $10^{-2}-10\mu$M for the sake of exploring ZOU. For the same reason, $A_0$ is also varied in the range 5.3-27.2 $\mu$M. }

\begin{tabular}{|l|l|l|}
\hline
\hline
 Symbol & Quantity & Estimated value \\ 
 \hline \hline 
 
 $ V $ & Cell Volume & $10^{-15}$ L \big(602.3 $\mu$M$^{-1}$\big) \\
 
 $B_0$ & CheB concentration & 0.28 $\mu$M \\ 
 
 $K_r$ & CheR dissociation constant & 0.39 $\mu$M \\ 
 
 $K_b$ & CheB dissociation constant & 0.54 $\mu$M \\ 
 
 $K_L$ & Ligand binding constant & 0.1 $\mu$M \\ 

 $\nu_r$ & Methylation rate & 0.75 $s^{-1}$ \\ 
 
 $\nu_b$ & Demethylation rate & 0.6 $s^{-1}$ \\ \hline \hline
\end{tabular}
\begin{flushleft}  
\end{flushleft}
\label{table1}
\end{table}

\subsection{Simulation results for $M=2$}

\begin{figure}[!h] 
\begin{center}
\includegraphics[width=14cm, angle = 0]{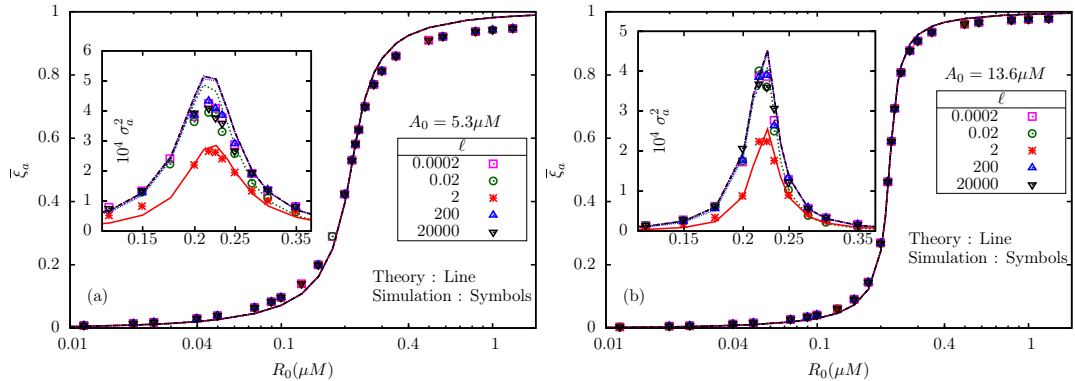}
\caption{The mean total active fraction $\overline{\xi}_a$ versus $R_0$ is independent of $\ell$; a manifestation of perfect adaptation as predicted by BL model. 
The figures show the results for two values of $A_0$, 5.3$\mu$M (a) and 13.6$\mu$M (b). In both the figures, the insets show the 
corresponding variances. Note that the maximum of the variance occurs at the same $R_0$ for various $\ell$.} 
\label{fig2}
\end{center}
\end{figure}

\begin{figure}[!h]
\begin{center}
 \includegraphics[width=8cm, angle = 0]{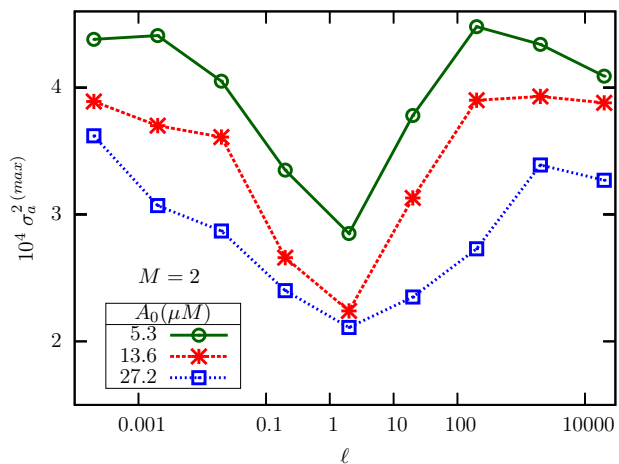}
\caption{The maximum variance, as determined from simulation data, is plotted against $\ell$ for three different $A_0$. 
Interestingly, the maximum is a non-monotonic function of $\ell$, and appears to have a minimum around $\ell=2$.} 
\label{fig3}
\end{center}
\end{figure}

\begin{figure}[!h]
\begin{center}
\includegraphics[width=8cm, angle = 0]{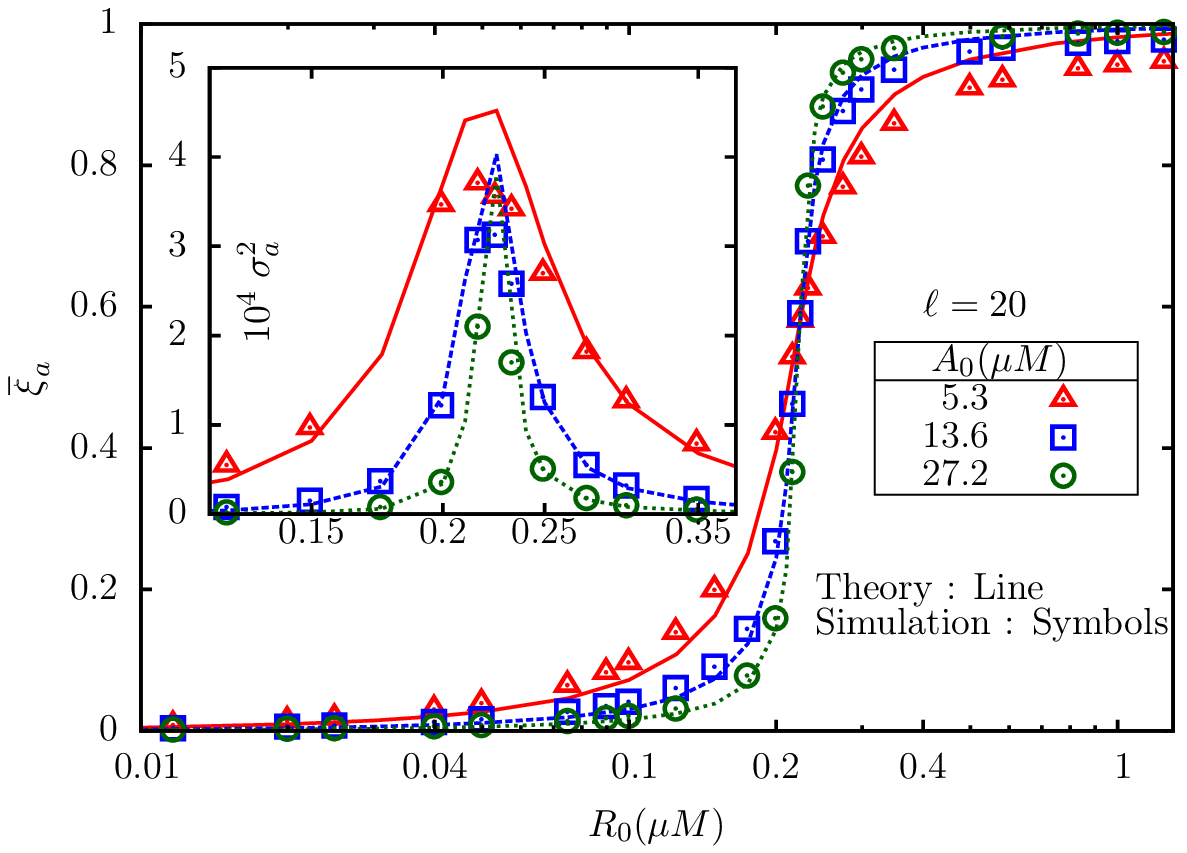}
\caption{The mean active fraction of receptors $\overline{\xi}_a$ plotted against $R_0$ for three different $A_0$ and fixed $\ell=20$. 
Inset: The corresponding variances.}
\label{fig4}
\end{center}
\end{figure}

For fixed $A_0$, the mean active fraction of receptors $\xi_a$ undergoes a sharp rise as $R_0$ is increased, as shown in Fig (\ref{fig2}). 
$\xi_a$ is independent of $\ell$, as is clear from the figure, and agrees with the analytical prediction (fixed point). 
The inset shows the variance in activity, which depends on $\ell$ in a non-monotonic manner (see Fig (\ref{fig3})). In all cases, the maximum 
of the variance occurs close to the point of the steepest rise in the mean. Note also that for $A_0=5.3\mu$M (Fig \ref{fig2}a), the rise is 
less steeper compared to $A_0=13.6\mu$M (Fig \ref{fig2}b); for same $\ell$, the variance is more in (a), but has a sharper peak in (b). 
For fixed $\ell$, and changing $A_0$, the mean shows the expected ultrasensitive rise as function of $R_0$ (Fig\ref{fig4}). 
For the set of parameters used 
here in Table \ref{table1}, the critical value for $R_0$ corresponding to $\alpha=1$ (\ref{alpha}) is $R_0^c=0.224\mu$M, which agrees 
with our observations in Fig\ref{fig4}. The variance (Fig \ref{fig4} inset) develops a sharper peak for larger $A_0$. However, note that, 
unlike our earlier observations\cite{bibZOU} in the GK model (M=1) with only fully inactive and fully active states, the peak value of the 
variance here does not seem to increase with $A_0$. 

\begin{figure}[!h]
\begin{center}
\includegraphics[width=14cm, angle = 0]{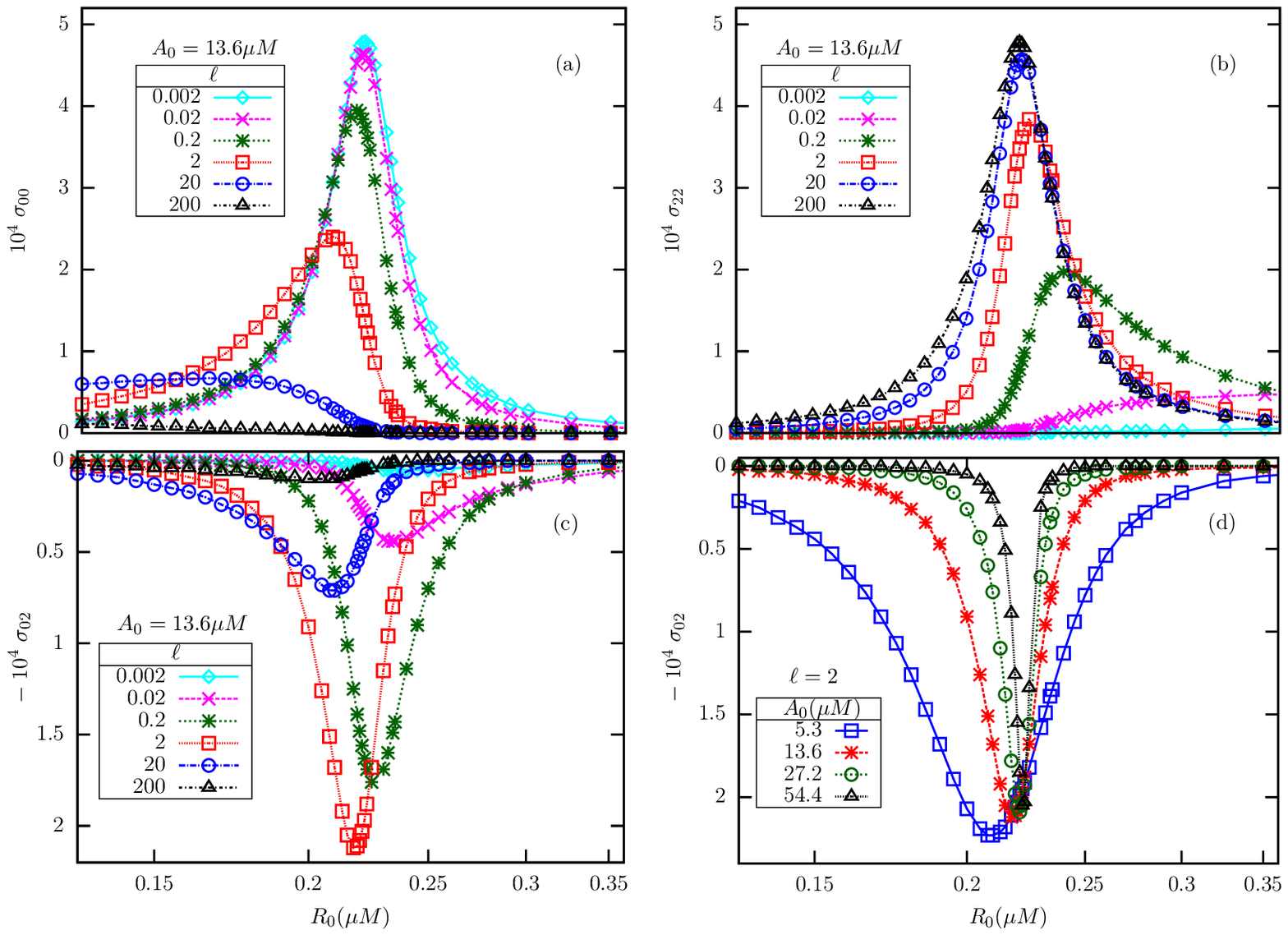}
\caption{The theoretical plots show the variances $\sigma_{00}$(a), $\sigma_{22}$(b) and the covariance $\sigma_{02}$(c and d). 
For fixed $A_0$, near the critical point, $\sigma_{00}$ is a decreasing function of $\ell$, while $\sigma_{22}$ is an increasing 
function of the same. By contrast, the covariance $\sigma_{02}$ (c) displays non-monotonic change with $\ell$ near the 
critical point, with a minimum reached near $\ell=2$ (compare with Fig \ref{fig3}). But $\sigma_{02}$ (d) varies monotonically 
with $A_0$ for fixed $\ell$.}
\label{fig5}
\end{center}
\end{figure}

Why would $\sigma_a^2$ show a non-monotonic variation with $\ell$? To find out, we studied the mathematical expressions for the individual 
variances $\sigma_{00}, \sigma_{22}$ and the covariance $\sigma_{02}$ as functions of $R_0$ and $\ell$, at fixed $A_0$. 
The plots, shown in Fig \ref{fig5}, indicate that the non-monotonicity originates from the covariance $\sigma_{02}$ 
between the fully inactive and fully active methylation levels. Both $\sigma_{00}$ (Fig \ref{fig5}a) and $\sigma_{22}$ (Fig \ref{fig5}b) have 
their peaks near $R_c$; but while the peak value of $\sigma_{00}$ is a decreasing function of $\ell$, that of $\sigma_{22}$ is an increasing 
function of $\ell$. $\sigma_{02}$ (Fig \ref{fig5}c and d) is negative throughout and has its minimum near $R_c$; however, in (c) the lowest 
value changes non-monotonically with $\ell$. From (\ref{ACTIVE}), it is seen that $\sigma_a^2$ is dominated by $\sigma_{00}$ for small $\ell$, 
while the large $\ell$ behavior is dominated by $\sigma_{22}$. In the intermediate $\ell$ regime, $\sigma_{02}$ also contributes significantly, 
and therefore the non-monotonic variation of its minimum value with $\ell$ affects the peak value of $\sigma_a^2$, which appears to be minimized near $\ell=2$.

\begin{figure}[!h]
\begin{center}
\includegraphics[width=14cm, angle = 0]{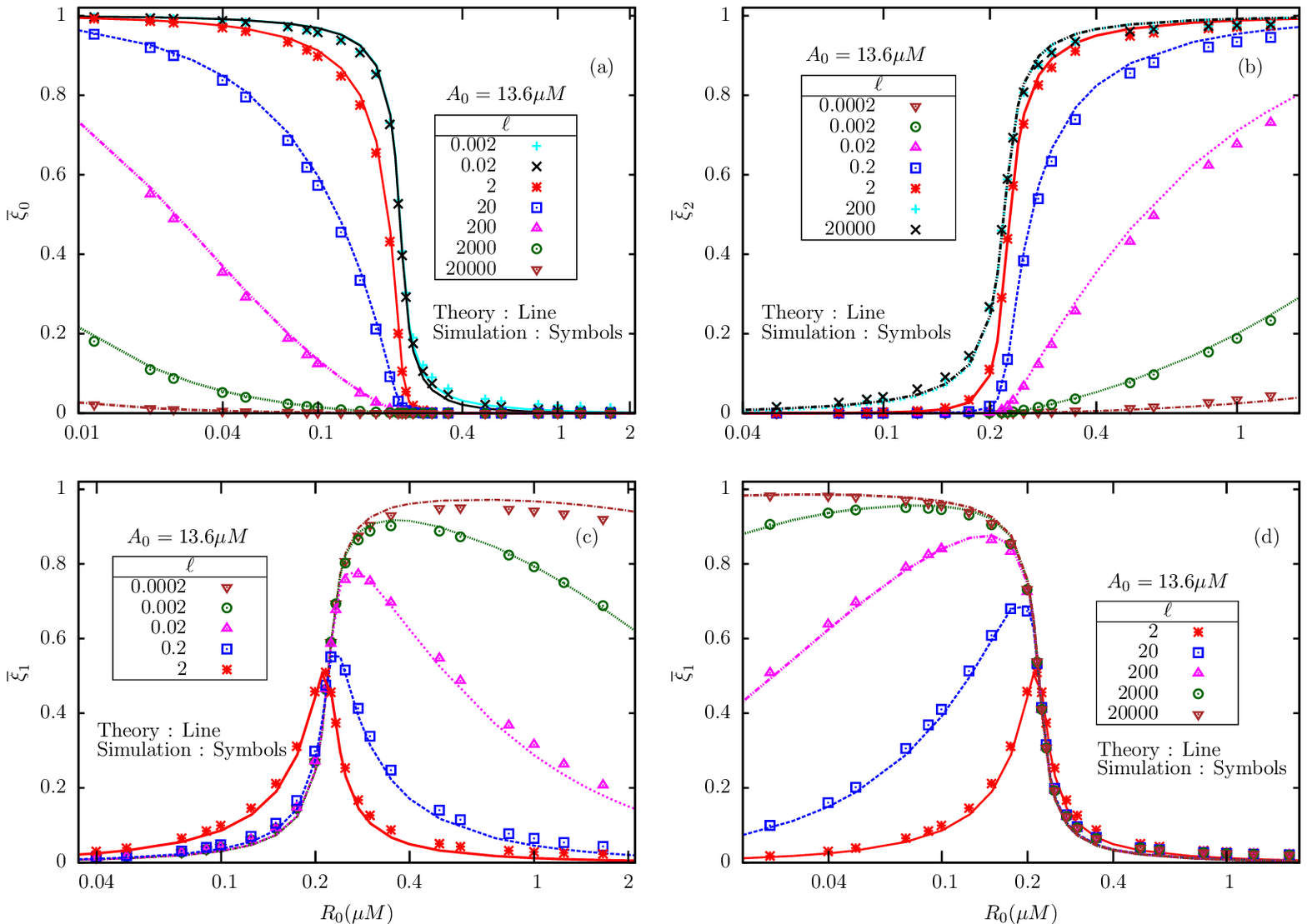}
\caption{The average fractions of receptors in different methylation levels (a) $\overline{\xi}_0$ (b) $\overline{\xi}_2$ (c) $\overline{\xi}_1$ at low ligand concentrations, and 
(d) $\overline{\xi}_1$ at high ligand concentrations,  as a function of an internal enzyme concentration ($R_0$), for fixed substrate concentration $A_0=13.6\mu$M.}
 \label{fig6}
 \end{center}
\end{figure}

\begin{figure}[!h]
\begin{center}
\includegraphics[width=8cm, angle = 0]{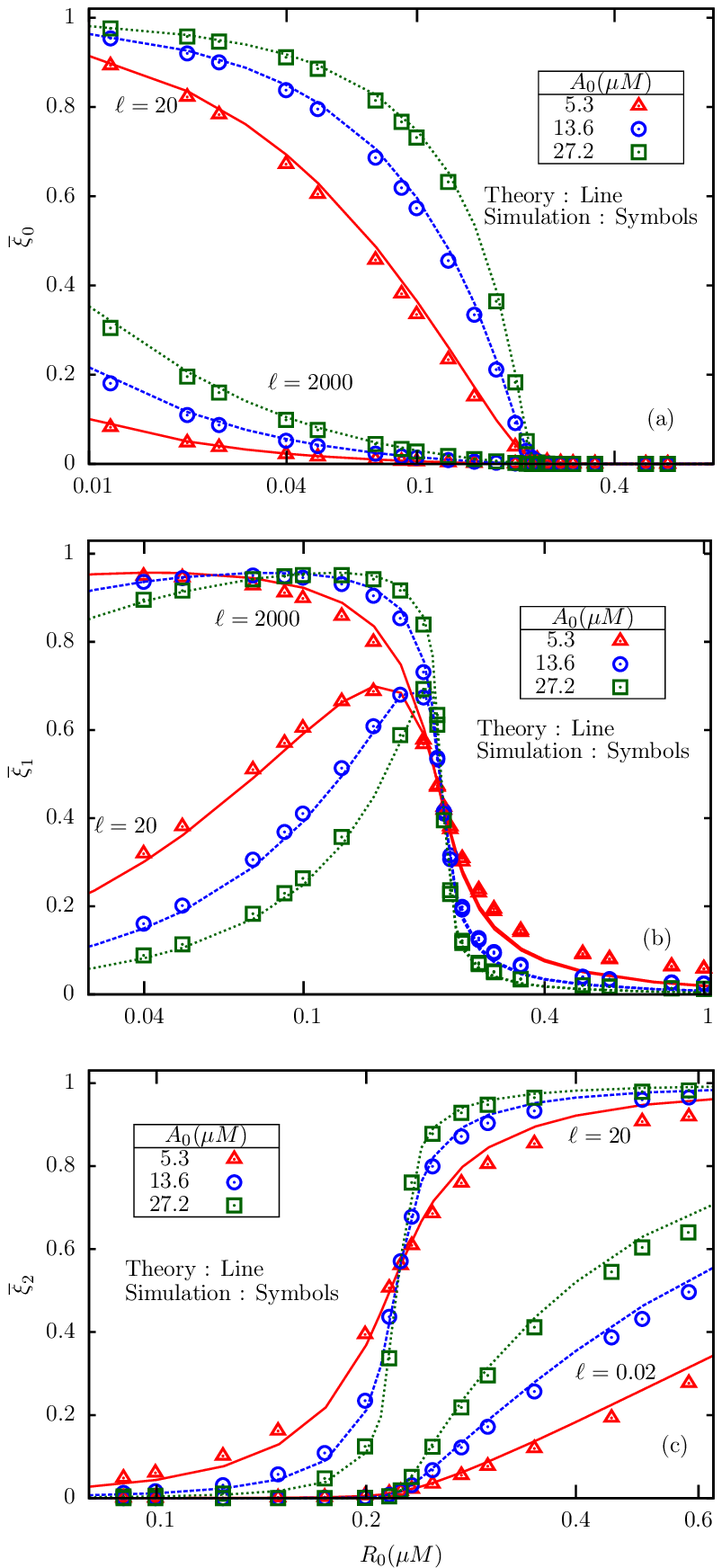}
\caption{A comparison of theoretical and simulation results for the mean fractions of receptors in three states, viz., $\overline{\xi}_0$, $\overline{\xi}_1$ and $\overline{\xi}_2$ 
plotted against $R_0$, for various substrate concentrations $A_0$, for two different values of $\ell$ (indicated in the figures).}
\label{fig7}
\end{center}
\end{figure}

Next, we try to understand how the receptors distribute themselves among different methylation levels
and the roles of $\ell$ and $A_0$ in this distribution. For fixed $A_0$ and very small $\ell$, the fraction of receptors 
in the lowest methylation level (always inactive) state shows an ultrasensitive transition from near-unity to near-zero across $R_c$. With 
increase in $\ell$, the transition becomes less sharper/smoother leading to $\xi_0$ being almost zero in the entire range of $R_0$ for very large $\ell$, as given in Fig (\ref{fig6}a). 
On the other hand, the fraction in the highest methylation level (always active) state is almost zero for very small $\ell$ and then increases in the upper-critical
regime with increase in $\ell$, finally leading to an ultrasensitive switch from near-zero to near-unity for very large $\ell$ Fig (\ref{fig6}b). 
But the fraction in the intermediate methylation level exhibits switch-like behaviors of opposite nature at very low and high $\ell$ values; 
while for intermediate $\ell$, it shows 
non-monotonic change with $R_0$ with a peak near $R_c$ Fig (\ref{fig6}c and d). With increase in $A_0$, the transitions in $\overline{\xi}_0$, $\overline{\xi}_1$ and $\overline{\xi}_2$ become sharper
(Fig \ref{fig7}), and agrees with the predictions made in (\ref{xi0020_sL}) and (\ref{xi0020_lL}). Therefore, in the $\ell\to 0$ as well 
as $\ell\to\infty$ regimes, the system effectively separates into two different GK-like two-state modules, with the population mostly shared 
between $m=0$ and $m=1$ states in the first case, and between $m=1$ and $m=2$ in the second case.

\subsection{Simulation results for $M>2$}

\begin{figure}[!h]
\begin{center}
\includegraphics[width=8cm, angle = 0]{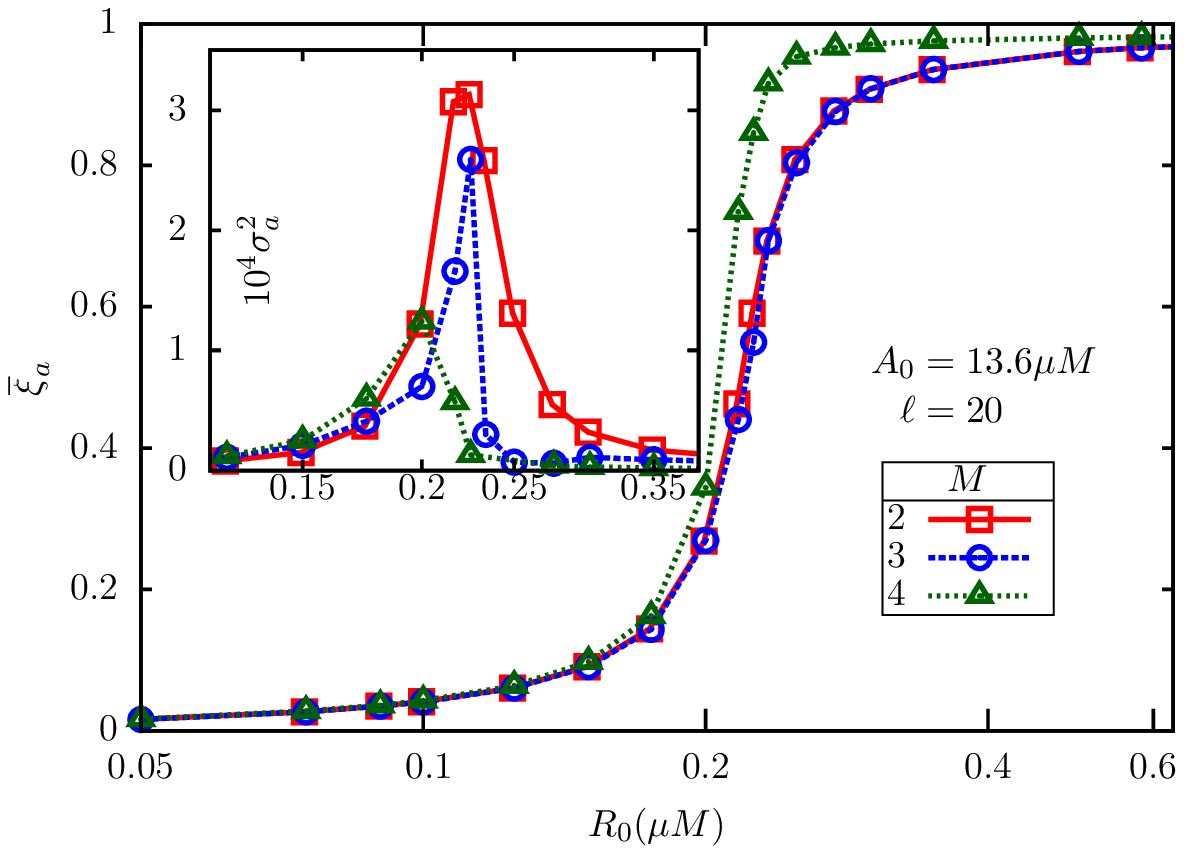}
\caption{The mean active fractions of receptors in BL models with $M=2,3,4$ in simulations, with substrate concentration fixed at 
$A_0=13.6\mu M$. The attractant concentration is $\ell=20$. The inset displays the corresponding variances. The numerical results 
establish that the qualitative behavior of the BL model, in particular, the ZOU transition, remains unchanged in presence of additional 
intermediate methylation levels.}
\label{fig8}
\end{center}
\end{figure}

Numerical simulations for models with more intermediate methylation levels, namely $M=3$ and $M=4$ were also carried out for a limited 
range of values of $\ell$ and $A_0$. Fig \ref{fig8} shows a comparison of the mean and variance in active fraction between models with $M=2,3$ 
and 4, for $\ell=20$ and $A_0=13.6\mu$M. For larger $M$, the mean shows sharper rise near $R_0=R_c$, but the peak value of the variance for 
$M=4$ is only about a third of the corresponding number for $M=2$. In general, models with larger number of intermediate states has smaller 
fluctuations in activity, presumably due to the dominant role of inter-state covariances. 

\begin{figure}[!h]
\begin{center}
\includegraphics[width=14cm, angle = 0]{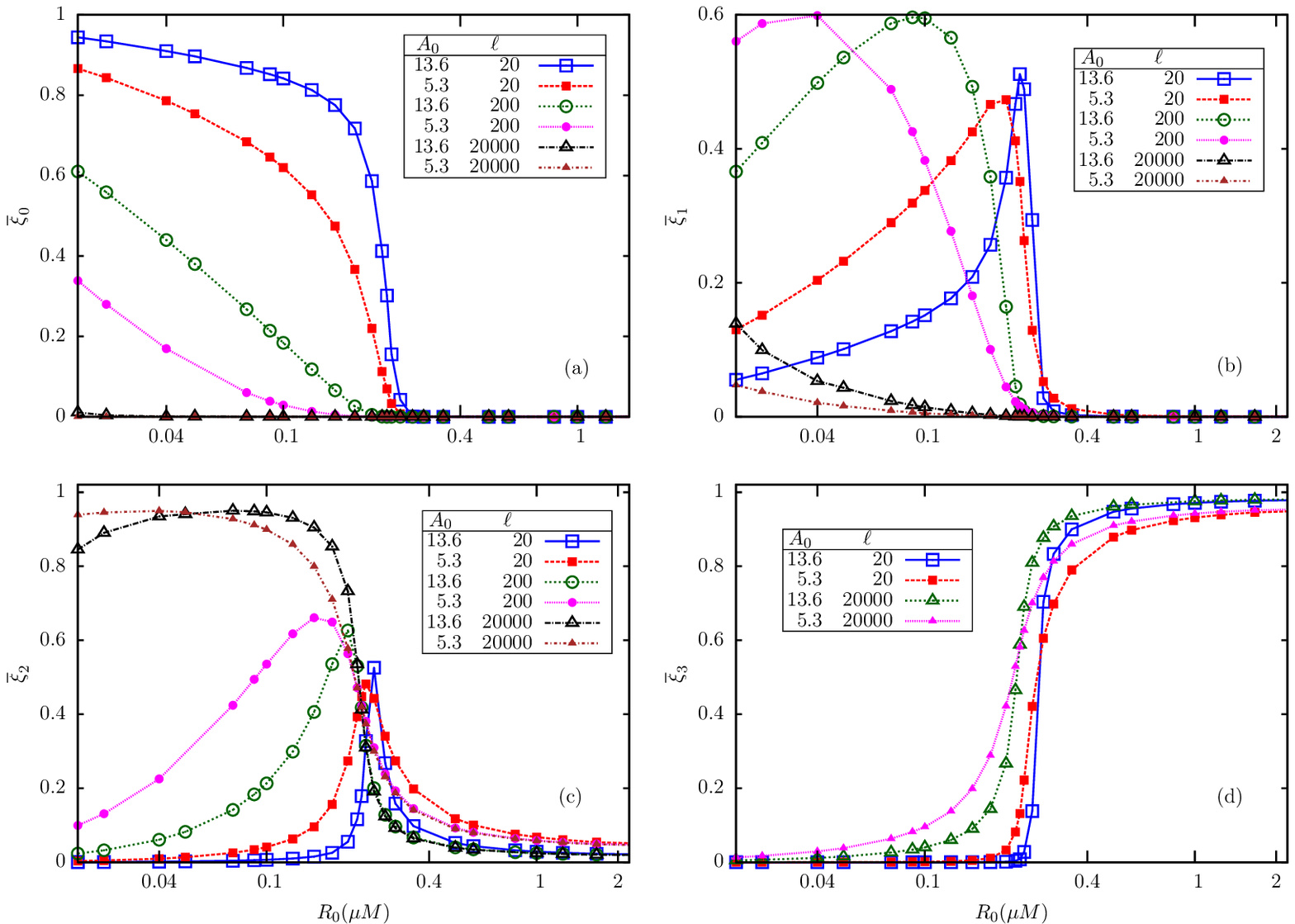}
\caption{Simulation results for the mean fractions of receptors in methylation states (a) $m=0$ (b) $m=1$ (c) $m=2$ and (d) $m=3$ in $M=3$ 
model, for two different substrate concentrations $A_0=5.3\mu$M and 13.6$\mu$M, and a few different values of $\ell$. }
\label{fig9}
\end{center}
\end{figure}

Next, we investigated whether the reduction to two independent two-state models which was found for $M=2$  works well for higher $M$ as well. In Fig \ref{fig9}a-d, we show, for $M=3$, 
the four mean populations $\overline{\xi}_0,\overline{\xi}_1,\overline{\xi}_2$ and $\overline{\xi}_3$ respectively for a limited range of $\ell$ and two different values of $A_0$. 
Similar to our observations in $M=2$ model, we see that the system effectively reduces to two different 2-state ($M=1$) models. For low $\ell$ ({\it e.g.} $\ell=20$), most of the 
receptors are in $m=0$ or $m=3$ states; in fact, $\overline{\xi}_0$ and $\overline{\xi}_3$ show near-ultrasensitive fall and rise respectively across $R_c$. For high $\ell$ ({\it e.g.} 
$\ell=20,000$), while both $\overline{\xi}_0$ 
and $\overline{\xi}_1$ become smaller as $A_0$ is increased, $\overline{\xi}_2$ and $\overline{\xi}_3$ become dominant; now, $\overline{\xi}_2$ falls abruptly as $R_0$ crosses $R_c$, 
while $\overline{\xi}_3$ rises in a similar way. Therefore, for $M=3$ as well, we observe that the system splits into two different GK modules in the 
limits $\ell\to 0$ and $\ell\to\infty$. 

The recent work by Pontius {\it et al.} \cite{bibPontius} also studied mean and fluctuations in receptor activity in {\it E.coli} by varying 
the enzyme ratio [CheR]/[CheB]. While their model includes more features of chemotaxis receptors like clustering, allosteric interactions 
and enzyme brachiation, it also suffers from a few drawbacks, in our opinion. (a)  In their analytical studies, the total methylation level 
in a cell is treated as the fundamental stochastic variable, whose dynamics is described by a phenomenological Langevin equation obtained by 
invoking the LNA; by contrast, our model has the receptor populations in various methylation levels as the basic 
variables, whose joint probability distribution follows a multivariate LFPE. (b) No direct quantitative comparisons 
between model predictions and stochastic simulations were done in their paper, while our model predictions are shown to agree well with 
simulations. (c) Attractant concentration was not 
included as a parameter in their model, while it is explicitly included in ours, and its effect on the ZOU transition has been explored in 
detail.

\subsection{Change in mean kinase activity in response to a change in attractant concentration}
Bacteria perform chemotaxis by responding to time-dependent changes in attractant/repellent concentration in its immediate environment. 
In this section, we explore the response characteristics of our 3-state model to a time-dependent change in attractant concentration, 
within linear response theory, applicable to weak perturbations. 
Let us consider a situation when a time-dependent change $\delta L (t)$ in the external ligand concentration is switched on in the 
extracellular environment of the bacterium at $t=0$, such that $L(t)\to L+\delta L(t)$ for $t\geq 0$, while $L(t)=L$ for $t<0$. Because ligand binding renders 
a receptor in the intermediate methylation level inactive, there is a change in the mean net activity at later times, which may be expressed in the general form 
 
\begin{equation} \label{chia}
\delta \overline{\xi_a}(t) =  \int_0^t dt' \chi_a (t-t')  \delta L (t')
\end{equation}
where $\chi_a(t)$ is the response function for mean receptor activity, which can be computed by  submitting the system to a (small) step-like change
in ligand concentration; $\delta L(t)=\delta L_s\Theta(t)$ where $\Theta(t)$ is the Heaviside theta-function and $\delta L_s$ is the size of the step. Let $\delta \overline{\xi_a}^{\rm (step)}(t)$ be the corresponding response. From (\ref{chia}), it follows that 
\begin{equation}
\chi_a(t)=\frac{1}{\delta L_s}\frac{d}{dt}\delta \overline{\xi}_a^{\rm (step)}(t)
\label{chia1}
\end{equation}

From (\ref{ACTIVE}), we find 
\begin{equation} \label{deltaXi}
\delta \overline{\xi}_a(t) = -\frac{K_L}{L+K_L}\delta \overline{\xi}_0(t) + \frac{L}{L+K_L}\delta \overline{\xi}_2 (t) 
 - \frac{K_L}{(L+K_L)^2}(1-\xi_0^*-\xi_2^*)\delta L (t),
\end{equation}
where $\delta \overline{\xi}_0(t)$ and $\delta \overline{\xi}_2(t)$ are the changes in $\overline{\xi}_0$ and $\overline{\xi}_2$, respectively, 
in response to a change $\delta L(t)$ in the ligand concentration, and are evaluated using (\ref{NEW}). The resulting equations have the form 
\begin{equation}\label{xit}
\frac{d}{dt}\delta\overline{\xi}_m=\sum_{n=0,2}\beta_{mn}\delta\overline{\xi}_n(t)+\gamma_m\delta L(t)~~~m=0,2
\end{equation}
where the coefficients $\beta_{mn}$ have been defined earlier, following (\ref{current}), while $\gamma_m=\partial v_m/\partial L|_{\boldsymbol{\xi}^*}$. 
Explicit expressions for both $\beta_{mn}$ and $\gamma_m$ are given in \ref{S2_Appendix}.

It is convenient to subject (\ref{deltaXi}) and (\ref{xit}) to Laplace transforms, after putting $\delta L(t)=\delta L_s\Theta(t)$, and use (\ref{chia1}) to compute the response function. For the Laplace transform of $\chi_a(t)$, we find 

 \begin{equation}\label{response1}
 \tilde{\chi}_a(s)= -\frac{K_L}{L+K_L} \tilde{\chi}_{_0}(s) + \frac{L}{L+K_L} \tilde{\chi}_{_2}(s) - \frac{K_L}{(L+K_L)^2} (1-\xi_0^*-\xi_2^*)
\end{equation}

where $\tilde{\chi}_{m}(s)$ are Laplace transforms of $\chi_m(t)$ ($m=0,2$) which are defined through the relations $\delta \overline{\xi}_m(t)=\int_{0}^{t}\chi_m(t-t^{\prime})\delta L(t^{\prime})dt^{\prime}$. The explicit expressions turn out to be 
\begin{eqnarray}\label{response2}
  \tilde{\chi}_{_0}(s) &=&  [ (s-\beta_{22})\gamma_0 + \beta_{02} \gamma_2 ] / d   \nonumber \\
  \tilde{\chi}_{_2} (s) &=&  [ \beta_{20} \gamma_0 +  (s-\beta_{00})\gamma_2 ]/ d, 
\end{eqnarray}
where $d = (s-\beta_{00})(s-\beta_{22}) - \beta_{02} \beta_{20}$. We have confirmed that the response function curve computed in (\ref{response1}) encloses zero area, i.e., $\tilde{\chi}_a(0)=0$, which is a manifestation of the perfect adaptation (regaining the pre-stimulus state within few seconds) property of the bacterium, reproduced in the BL model (in our notation, this property implies lack of dependence of $\overline{\xi}_a$ on $\ell$ in steady state, recall Fig (\ref{fig2})). 
In \ref{S3_Appendix}, we use the results in (\ref{response1}) and (\ref{response2}) to derive the chemotactic response function, 
which determines the fractional change in clockwise bias of the flagellar motor in response to temporal variations in attractant 
concentration, and consequently the chemotactic drift.

\section{Discussion}
The subject of the present study is the BL model of conformational dynamics of chemotaxis receptors in {\it Escherichia coli}. 
According to this model, a receptor undergoes stochastic switching between its active and inactive states, depending on its level of 
methylation and ligandation. The methylation and demethylation of receptors is carried out by two enzymes, viz. methyltransferase CheR
and methylesterase CheB, acting antagonistically; 
the overall structure of the circuit is similar to the two-state ({\it e.g.} active/inactive) covalent modification scheme studied originally 
by GK \cite{bibGK}. This system is characterized by ZOU, a switch-like transition 
between all-inactive to all-active phases, in the limit of infinitely large substrate concentration. In the present paper, we investigated 
ZOU in the BL model within a stochastic formulation where the number $N$ of receptors is treated finite. For analytical tractability,
we first study a model with a single intermediate partially active state sandwiched between the fully inactive and fully active states. We 
show rigorously that the mean activity shows ultrasensitive response to increase in [CheR] (at fixed [CheB]) in the limit $N\to\infty$, 
independent of the attractant concentration $L$. At the same time, its variance $\sigma_a^2$ is a non-monotonic function of [CheR] and $L$, 
as deduced from linear-noise approximation (LNA) and confirmed in numerical simulations. Interestingly, for very low and very high $L$, 
the system effectively reduces to two different ``two-state'' modules, akin to GK switches. The analysis is extended to models with more intermediate methylation levels ({\it E.coli} receptor has 3) via numerical simulations. By and large, qualitatively similar behavior is found in all the cases. 
For {\it E.coli} parameters, we find that $\sigma_a\simeq 10^{-2}$ in the vicinity of the critical point of ZOU, for a large range of $L$, 
spanning almost 8 orders of magnitude. This estimate agrees with another recent computational study where explicit receptor clustering 
was included, and receptor activation-inactivation dynamics was stimulated using an equilibrium MWC model \cite{bibPontius}. 
In \ref{S3_Appendix}, we show that this is also consistent with experimental measurements of [CheYp] fluctuations. 

Biochemical noise is likely to be a crucial factor in the motility of {\it E.coli} for the following reason. The transition from run to tumble 
mode of a flagellar motor in {\it E.coli} is brought about by a switch in its direction of rotation, from counter-clockwise (CCW) to clockwise 
(CW). The latter switching is known to occur in an ultrasensitive fashion as a function of [CheYp], the concentration of phosphorylated CheY, 
which is the response regulator cytoplasmic protein (the phosphorylation being done by active receptors). 
The ultrasensitive response of flagellar motor for the CW bias with the variation in the concentration of CheYp, 
was first reported by Cluzel {\it et al.} \cite{bibCluzel} by performing {\it in vivo} experiments in single cells in absence of 
stimulant (attractant/repellent). 
The authors observed cell to cell variability in the amount of CheYp which
was distributed around a mean value (ranging from $0.8$ to $6 \mu M$) even for a given
concentration of the inducer isopropyl $\beta$-D-thiogalactoside, in short IPTG (used to express CHEY-GFP in PS2001 {\it E. coli} strain). CW bias
in various cells, pre-induced with different inducer levels, collapsed onto a single sigmoid curve
when plotted against CheYp. The measured CW bias in the range $0.1$ to $0.9$ could be fitted to a Hill plot with Hill coefficient $H\sim
10.3\pm 1.1$, with the sensitive part being less than $3\mu$M in width. The ability of the cell to execute runs and tumbles in 
succession depends on placing the intracellular [CheYp] somewhat precisely in this window.

Cluzel {\it et al.} proposed that an additional molecular mechanism such as CheZ acting as the [CheYp] regulator can help in retaining the
concentration of the latter in the narrow sensitive window of operation. Later, Yuan {\it et al.} \cite{bibYuan_Nature} found 
from their experiments with {\it cheRcheB} cells that the flagellar motor can partially adapt to changes in [CheYp] over a time 
scale of several minutes, and thereby laterally shift its operating range by about 0.5$\mu$M.  Any likelihood of partial adaptation 
due to dynamic localization of CheZ (suggested by Lipkow \cite{bibLipkow}) was ruled out by working with {\it cheRcheBcheZ} 
cells and explicitly showing similar results in both the cases. In 2013, Yuan and Berg \cite{bibYuan_JMB} carried out a new set of 
experiments with pre-adapted motors, which showed that the response curve in this case is much steeper, compared to adapted motors 
(as in \cite{bibCluzel}) with Hill coefficients $\sim 16.5 \pm 1.1$ and $20.7 \pm 1.6$ for CW biases of $0.8 \pm 0.1$ and $0.5 \pm 0.1$ 
respectively. Because of the higher Hill coefficient, the sensitive window of [CheYp] is reduced to about 1$\mu$M in this case, with 
center around 3$\mu$M. 

What is the impact of receptor-level noise on the clockwise bias of a single cell ?  It is clear that spontaneous fluctuations in receptor 
activity will produce corresponding changes in [CheYp], which would, in turn, affect the CW bias of the motor(s).  
In \ref{S4_Appendix} we estimate that the maximum standard deviation in clockwise bias (arising from spontaneous 
fluctuations in [CheYp]) $\delta P_{CW}$ lies in the range 0.15-0.36 for Hill coefficient $H=20$ \cite{bibYuan_JMB}, with the lower 
value corresponding to mean CW bias $\sim$ 0.5 and the upper value for mean CW bias $\sim$ 0.1. It therefore appears that 
intrinsic noise could rescue a cell with mean [CheYp] outside the optimal range from being forced to "run forever" without tumbling. 
The motor-level adaptation modifies these estimates a little, but our detailed analysis presented in \ref{S5_Appendix} shows that 
this effect is negligible when [CheYp] lies in the range $ 2.5 \mu M - 4 \mu M $, which more than covers the sensitive window. 
Outside this range, motor-level adaptation increases the standard deviation in CW bias, (see figure in \ref{S5_Appendix}). We conclude, therefore, 
that intrinsic noise plays an important role in the run and tumble behavior of {\it E.coli}, and by extension, in chemotactic drift as well. 
A detailed analysis of the latter is presently being carried out and will be reported in the near future.

Being a study focused primarily on exploring the occurrence of ZOU in the BL model, we have not included all the features of 
signal transduction in {\it E.coli} here, even at the level of receptor dynamics. In particular, clustering of receptors and allosteric 
interactions among them have not been included here, but have been studied recently by other authors\cite{bibPontius, bibEndres, bibHansen}. 
In \cite{bibPontius}, the authors show via numerical simulations that receptor clustering and enzyme localization make the mean activity a less sensitive function of [CheR], and thus makes the network robust to variations in protein numbers. At the same time, enzyme localization also leads to larger 
fluctuations in activity, with almost 10-fold increase in variance. Intrinsic biochemical noise arising from spontaneous fluctuations in 
receptor activity has been shown to be important in explaining the kinetics of flagellar motor-switching\cite{bibPark_Nature, bibPark_BiophysJ}. Using a set of parameters extracted from the literature, we show 
in \ref{S3_Appendix} that the variance in activity measured in our simulations (as well as predicted using LNA) appear to be sufficiently 
large in magnitude to generate the [CheYp] fluctuations observed in experiments\cite{bibPark_Nature}. It is therefore likely that receptor clustering and enzyme localization may not be necessary to enhance the noise levels, although it could be important in other aspects of signaling, e.g. signal amplification. 


\section{Appendices}


\subsection{Intra and inter-module dynamics in the 3-state receptor-level biochemical reactions}
\label{S1_Appendix}

To evaluate the number of receptors in each of the configurations, we employ sQSSA. 
The essence of sQSSA lies in the assumption that all the modules are weakly coupled ($\nu_r$ and $\nu_b$ assumed to be very small 
compared to other rates involved in the system) and hence detailed balance exists within each module. Under this condition, the mean numbers 
of intermediate complexes can be expressed in terms of their unbound counterparts as 

\begin{equation}
\overline{\tilde{x}_0}=\frac{R_f}{K_r}\overline{x_0}~~;~~\overline{\tilde{x}_1^i}=\frac{R_f}{K_r} \overline{x_1^i}~~;~~
\overline{\tilde{x}_1^a}=\frac{B_f}{K_b} \overline{x_1^a}~~;~~\overline{\tilde{x}_2}=\frac{B_f}{K_b} \overline{x_2}~;~~
\overline{x_1^i}=\frac{L}{K_L} \overline{x_1^a}.
\end{equation}

where $R_f$ and $B_f$ are the mean concentrations of free/unbound R and B enzymes respectively. After using the module-wise normalization 
conditions $x_0+\tilde{x}_0=\xi_0$, $x_1^i+\tilde{x}_1^i=\xi_1^i$, $x_1^a+\tilde{x}_1^a=\xi_1^a$ and $x_2+\tilde{x}_2=\xi_2$, we find 

\begin{eqnarray} \label{intra}
 \overline{\tilde{x}_0}(\boldsymbol{\xi}) &=& \frac{R_f}{R_f + K_r}  \xi_0  \nonumber \\
 \overline{\tilde{x}_1^i}(\boldsymbol{\xi}) &=& \frac{R_f}{R_f + K_r}  \overline{\xi_1^i} \nonumber \\
 \overline{\tilde{x}_1^a}(\boldsymbol{\xi}) &=& \frac{B_f}{B_f + K_b}  \overline{\xi_1^a} \nonumber \\
 \overline{\tilde{x}_2}(\boldsymbol{\xi}) &=& \frac{B_f}{B_f + K_b}  \xi_2 \nonumber \\
 \overline{\xi_1^i} &=& \frac{L}{K_L}\overline{\xi_1^a}.
\end{eqnarray}

In the last equation, we have used the assumption of infinitely fast ligand binding and dissociation kinetics. Inter-module equilibrium is expressed through the following flux-balance relations which determine the fixed point $\boldsymbol{\xi}^*$:

\begin{equation} \label{inter}
 \nu_r \overline{\tilde{x}_0}(\boldsymbol{\xi^*})-\nu_b \overline{\tilde{x}_1^a}(\boldsymbol{\xi}^*)=0=\nu_r\overline{\tilde{x}_1^i}(\boldsymbol{\xi}^*) -\nu_b\overline{\tilde{x}_2}(\boldsymbol{\xi}^*). 
\end{equation}

The free enzyme concentrations $R_f$ and $B_f$ are determined by the normalization conditions

\begin{eqnarray} \label{RoBo}
R_f(\boldsymbol{\xi}) + A_0 [\overline{\tilde{x_0}}+\overline{\tilde{x}_1^i}] = R_0 \nonumber \\
B_f(\boldsymbol{\xi}) + A_0 [\overline{\tilde{x_2}}+\overline{\tilde{x}_1^a}] = B_0
 \end{eqnarray}
 
 Using (\ref{intra}) in (\ref{RoBo}) leads to (\ref{RB}). \newline

\noindent {\bf Perfect adaptation (insensitivity of average activity to {\it L}):} \newline

Although the fixed point $\boldsymbol\xi^*$ depends on $L$, the mean total active fraction does not; this can be easily shown as follows. 
The equations (\ref{inter}) for intermodule dynamics, after summing the left hand side give

\begin{equation} \label{interSum}
\nu_r (\overline{\tilde{x}_0}+\overline{\tilde{x}_1^i}) = \nu_b (\overline{\tilde{x}_1^a}+\overline{\tilde{x}_2})
\end{equation}

After substituting the equations for intra-module dynamics (\ref{intra}) (in the limit $R_f \ll K_r$, $B_f \ll K_b$) and the expression for 
$\xi_a$ in terms of $\xi_m$ (\ref{ACTIVE}) in (\ref{interSum}), the following quadratic equation for $\xi_a^*$, independent of $L$, is obtained:

\begin{equation}
 \xi_a^{*2}(\nu_bB_0-\nu_rR_0) + \xi_a^*[\nu_rR_0(A_0-K_b)-\nu_bB_0(K_r+A_0)] + \nu_rR_0A_0K_b = 0.
\end{equation}

The solution to the above equation is 

\begin{eqnarray}
 \xi_a^*&=&\frac{\nu_bB_0(A_0+K_r)-\nu_rR_0(A_0-K_b)}{2(\nu_bB_0-\nu_rR_0)} \nonumber \\
 &+&\frac{\sqrt{[\nu_bB_0(A_0+K_r)-\nu_rR_0(A_0-K_b)]^2-4\nu_rR_0A_0K_b(\nu_bB_0-\nu_rR_0)}}
 {2(\nu_bB_0-\nu_rR_0)}
\end{eqnarray}

\subsection{Evaluation of various parameters}
\label{S2_Appendix}

The expressions of $\beta_{mn}$ can be obtained by differentiating (\ref{V}).
Note that the expressions for $v_m$ contain $R_f$ and $B_f$ which are in turn functions of $\xi_m$. Therefore, 

\begin{equation}
 \beta_{mn} = \Bigg( \frac{\partial v_m}{\partial \xi_n} + \frac{\partial v_m}{\partial R_f} \frac{\partial R_f}{\partial \xi_n}
 + \frac{\partial v_m}{\partial B_f} \frac{\partial B_f}{\partial \xi_n} \Bigg)_{\boldsymbol{\xi^*}},
\end{equation}

where
\begin{eqnarray}
 \frac{\partial v_0}{\partial \xi_0} &=& - \Big( \frac{\nu_b B_f}{K_b} \frac{K_L}{L+K_L} +  \frac{\nu_r R_f}{K_r} \Big) \nonumber \\
 \frac{\partial v_0}{\partial \xi_2} &=& - \frac{\nu_b B_f}{K_b} \frac{K_L}{L+K_L}  \nonumber \\
 \frac{\partial v_2}{\partial \xi_0} &=& - \frac{\nu_r R_f}{K_r} \frac{L}{L+K_L} \nonumber \\
 \frac{\partial v_2}{\partial \xi_2} &=& - \Big( \frac{\nu_r R_f}{K_r} \frac{L}{L+K_L} +  \frac{\nu_b B_f}{K_b} \Big) \nonumber \\
 \end{eqnarray}
\begin{eqnarray}
 \frac{\partial v_0}{\partial R_f} &=& - \frac{\nu_r}{K_r}\xi_0 \nonumber \\
 \frac{\partial v_0}{\partial B_f} &=& \frac{\nu_b}{K_b}\frac{K_L}{L+K_L}(1-\xi_0-\xi_2) \nonumber \\
 \frac{\partial v_2}{\partial R_f} &=& \frac{\nu_r}{K_r}\frac{L}{L+K_L}(1-\xi_0-\xi_2) \nonumber \\
 \frac{\partial v_2}{\partial B_f} &=& - \frac{\nu_b}{K_b}\xi_2^* 
 \end{eqnarray}
\begin{eqnarray}
 \frac{\partial R_f}{\partial \xi_0} &=& - \frac{R_0 K_r (L+K_L) A_0 K_L}{ \Big[ K_r(L+K_L) + A_0 \big( K_L \xi_0 + L (1-\xi_2) \big) \Big]^2} \nonumber \\
 \frac{\partial R_f}{\partial \xi_2} &=& \frac{R_0 K_r (L+K_L) A_0 L}{ \Big[ K_r(L+K_L) + A_0 \big( K_L \xi_0 + L (1-\xi_2) \big) \Big]^2 } \nonumber \\
 \frac{\partial B_f}{\partial \xi_0} &=&  \frac{B_0 K_b (L+K_L) A_0 K_L}{ \Big[ K_b(L+K_L) + A_0 \big( K_L (1-\xi_0) + L \xi_2 \big) \Big]^2} \nonumber \\
 \frac{\partial B_f}{\partial \xi_2} &=& - \frac{B_0 K_b (L+K_L) A_0 L}{ \Big[ K_b(L+K_L) + A_0 \big( K_L (1-\xi_0) + L \xi_2 \big) \Big]^2 }
 \end{eqnarray}
To evaluate the linear response function, we have to consider the derivative $\gamma_m$, the explicit expression of which is 
\begin{equation}\label{gamma}
\gamma_m=\bigg(\frac{\partial v_m}{\partial L}+\frac{\partial v_m}{\partial R_f}\frac{\partial R_f}{\partial L} 
+ \frac{\partial v_m}{\partial B_f}\frac{\partial B_f}{\partial L} \bigg)_{{\boldsymbol \xi}^*}.
\end{equation}
The relevant quantities involving derivatives with respect to $L$ are listed below:
\begin{eqnarray}
  \frac{\partial v_0}{\partial L} &=& - \frac{\nu_b B_f}{K_b} \frac{K_L}{(L+K_L)^2} (1-\xi_0-\xi_2) \nonumber \\
  \frac{\partial v_2}{\partial L} &=& \frac{\nu_r R_f}{K_r} \frac{K_L}{(L+K_L)^2} (1-\xi_0-\xi_2)
\end{eqnarray}
\begin{eqnarray}
\frac{\partial R_f}{\partial L} &=& - \frac{R_0 K_r A_0 K_L (1-\xi_0-\xi_2)}{ \Big[ K_r(L+K_L) + A_0 \big( K_L \xi_0 + L (1-\xi_2) \big) \Big]^2 } \nonumber \\
\frac{\partial B_f}{\partial L} &=& \frac{B_0 K_b A_0 K_L (1-\xi_0-\xi_2)}{ \Big[ K_b(L+K_L) + A_0 \big( K_L (1-\xi_0) + L \xi_2 \big) \Big]^2 }
\end{eqnarray}
In the computation of the coefficients $\beta_{mn}$ and $\gamma_m$, all the above mentioned quantities are to be evaluated at the 
fixed point $(\boldsymbol{\xi}^*,R_f(\boldsymbol{\xi}^*),B_f(\boldsymbol{\xi}^*)$ and it has been assumed that $R_f \ll K_r$ and 
$B_f \ll K_b$ .

\subsection{Chemotactic Response Function}
\label{S3_Appendix}

In {\it E.coli}, the receptor complex acts as the processing unit for any input signal (e.g. changes in the extracellular environment of the bacterium)
while the flagellar motor, which controls the switching kinetics  of the flagella,  regulate the output (run/tumble motion). While the CCW mode of rotation of flagella corresponds to straight motion of the bacterium, the CW mode corresponds to tumbles. 
These two key constituents of the signaling network are connected by the response regulator CheY. Attractant binding to the receptors 
results in suppression of the activity of the latter, hence leads to less phosphorylation of CheY. Phosphorylated CheY (CheYp) binds to flagellar motors and enhances the rate of CCW$\to$CW switching, hence a temporal increase in attractant concentration as sensed by the bacterium would result in elongated run durations when swimming in favorable directions. 
The dynamics of phosphorylated CheY may be described by the equation
\begin{equation}\label{s41}
\frac{dY(t)}{dt}=a_{Y}[Y_0-Y(t)]A_0\xi_a(t)-\lambda_{Y}Y(t)
\end{equation}
where $a_Y$ is rate for binding of (non-phosphorylated) CheY to an active receptor, which leads to its phosphorylation, $Y_0$ is the total 
concentration of CheY (henceforth assumed much larger than $Y(t)$) and $\lambda_Y$ is the rate of degradation of CheYp (due to the 
dephosphatase CheZ). The steady state mean concentration of CheYp can be obtained from (\ref{s41}) by equating left hand side to $0$,
\begin{equation}\label{Y*}
 Y^* = \frac{a_Y Y_0 A_0}{\lambda_Y} \xi_a^* .
\end{equation}
In response to a change in attractant concentration $L\to L+\delta L(t)$ for $t\geq 0$, the mean [CheYp] undergoes a shift 
$Y^*\to Y^*+\delta Y(t)$, with $\delta Y(t)$ given by 
\begin{equation} \label{Y_soln}
 \delta Y(t) = a_Y Y_0 A_0 \int_0^t e^{- \lambda_Y(t-t')} \delta \overline{\xi_a}(t')dt' .
\end{equation}

\begin{figure}[!h]
\begin{center}
\includegraphics[width=14cm, angle = 0]{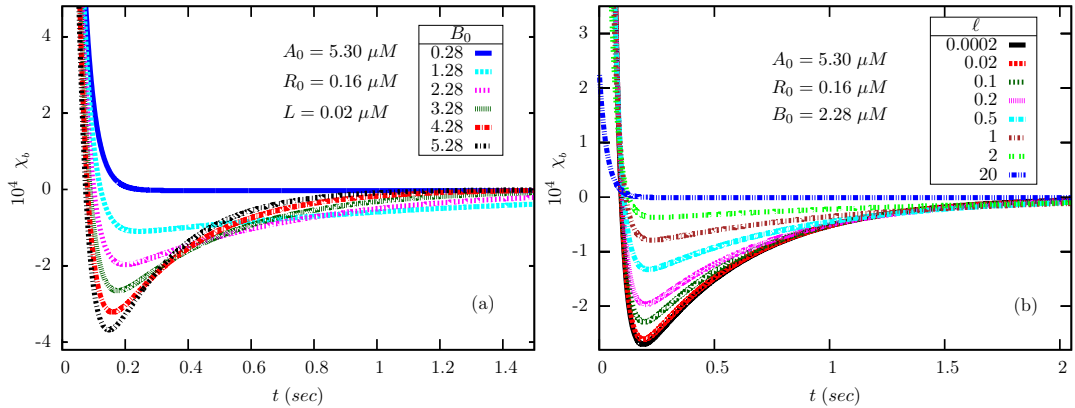}
\caption{The bilobed chemotactic response function $\chi_{b}(t)$ (\ref{CHIb_final}) as a function of time obtained for various values of 
(a) methylesterase CheB ($B_0$) and (b) ligand ($\ell=L/K_L$) concentrations. In (a), we have fixed $\ell=0.2$. All other (fixed) 
parameters take the values mentioned in Table\ref{table1}. 
 We have also used $\lambda_Y = 30 s^{-1} $ \cite{bibReneaux,bibRao_JCP, bibKollmann}, $H=20$ \cite{bibYuan_JMB} and $P_{\rm CW}(Y^*)=0.5$.} 
 \label{figS4_1}
\end{center}
\end{figure}

The {\it clockwise bias}, i.e., the probability $P_{\rm{CW}}$ for a flagellar motor to be in clockwise-spinning state (corresponding to tumble) 
has been found in experiments to be an ultrasensitive function of $Y$\cite{bibYuan_JMB, bibMello, bibCluzel}: 
\begin{equation}\label{s44}
P_{\rm{CW}}(Y)\approx \frac{Y^H}{Y^H+K_Y^H},
\end{equation}
where $H$ is the Hill coefficient symbolizing the steepness of the transition. It follows that fractional changes in the CW bias $P_{\rm{CW}}$ and $Y$ are related as 
\begin{equation}\label{deltaR}
 \frac{\delta P_{\rm{CW}} (t)}{P_{\rm{CW}}} = H (1-P_{\rm{CW}})\frac{\delta Y (t)}{Y}=\int_{0}^{t}\chi_b(t-t^{\prime})\delta L(t^{\prime})dt^{\prime}, 
\end{equation}
where, in the second part, we have defined the response function $\chi_b(t)$ for the bias. Using (\ref{Y_soln}) and (\ref{Y*}) in (\ref{deltaR}), 
it follows that, in steady state, $\chi_b(t)$ is related to $\chi_a(t)$ through 

\begin{equation} \label{CHI_b}
\chi_{b} (t) = \frac{H\lambda_Y}{ \xi_a^*}\bigg [1-P_{\rm{CW}}(Y^*)\bigg]\int_{0}^{t}e^{-\lambda_{Y}(t-t^{\prime})}\chi_a(t^{\prime})dt^{\prime}
\end{equation}
From (\ref{CHI_b}), it follows that the Laplace transforms of $\chi_b(t)$ and $\chi_a(t)$ are related linearly: $\tilde{\chi_b}(s)\propto \tilde{\chi_a}(s)/(s+\lambda_Y)$, hence $\tilde{\chi_b}(0)=0$, i.e., the response curve for the bias too encloses zero area\cite{bibSegall,bibReneaux}. After using (\ref{response1}) and (\ref{response2}) in (\ref{CHI_b}), and performing the inversion, we arrive at the following multi-exponential form for $\chi_b(t)$: 

\begin{eqnarray}\label{CHIb_final}
 \chi_{b} (t) = && \frac{H \lambda_Y}{\xi_a^*}\bigg [1-P_{\rm{CW}}(Y^*)\bigg] \Bigg[ e^{-\lambda_Y t}\Big( -\frac{K_L X_0}{L+K_L} + \frac{L X_2}{L+K_L}
 - \frac{K_L (1-\xi_0^*-\xi_2^*)}{(L+K_L)^2} \Big)  \nonumber \\
 && + e^{-At} \Big( -\frac{K_L Y_0}{L+K_L} + \frac{L Y_2}{L+K_L} \Big)
 + e^{-Bt} \Big( -\frac{K_L Z_0}{L+K_L} + \frac{L Z_2}{L+K_L} \Big) \Bigg]
\end{eqnarray}
The rate constants $A$ and $B$, as well as the coefficients $X_m$, $Y_m$ and $Z_m (m=0,2)$ are expressed in terms of $\beta_{mn}$ and $\gamma_{m}$ as follows: 
\begin{eqnarray}
 A &=& -\frac{1}{2}\Big[ (\beta_{00}+\beta_{22})+\sqrt{(\beta_{00}-\beta_{22})^{2}-4\beta_{02}\beta_{20}} \Big] \nonumber \\
 B &=& -\frac{1}{2}\Big[ (\beta_{00}+\beta_{22})-\sqrt{(\beta_{00}-\beta_{22})^{2}-4\beta_{02}\beta_{20}} \Big]
\end{eqnarray}
\begin{eqnarray}
 X_0 & = &\frac{-(\lambda_Y+\beta_{22})\gamma_0 + \beta_{02}\gamma_2 }{(\lambda_Y+A)(\lambda_Y+B)}~~;~~X_2  =  \frac{\beta_{20}\gamma_0 - (\lambda_Y+\beta_{00})\gamma_2 }{(\lambda_Y+A)(\lambda_Y+B)} \nonumber \\
 Y_0 & = & \frac{(A-\beta_{22})\gamma_0 + \beta_{02}\gamma_2 }{(\lambda_Y+A)(A-B)}~~;~~Y_2  =  \frac{\beta_{20}\gamma_0 - (A-\beta_{00})\gamma_2 }{(\lambda_Y+A)(A-B)} \nonumber \\
 Z_0 & = & \frac{(B-\beta_{22})\gamma_0 + \beta_{02}\gamma_2 }{(\lambda_Y+B)(B-A)}~~;~~Z_2  =  \frac{\beta_{20}\gamma_0 + (B-\beta_{00})\gamma_2 }{(\lambda_Y+B)(B-A)}
\end{eqnarray}
In Fig.\ref{figS4_1}, we show plots of the mathematical expression for $\chi_b(t)$ obtained in (\ref{CHIb_final}), 
(a) by varying $B_0$ and (b) varying $\ell$.  The curves closely resemble the experimentally measured 
responses to short-lived stimuli \cite{bibSegall, bibMasson}, for somewhat large values of $B_0$, in agreement with 
earlier results in a mean-field BL model\cite{bibReneaux}. In (a), with increase in $B_0$, there is 
an overall reduction in time scales and an increase in the depth of the negative lobe. In (b), with increase in $\ell$, 
the depth of the negative lobe decreases wih no change in the time scales. For $\ell\gg 1$, the negative lobe effectively disappears. 

\subsection{ On the role of biochemical noise in the run and tumble motion of {\it E.coli} }
\label{S4_Appendix}

From (\ref{s44}), it follows that a fluctuation in [CheYp] equal to its standard deviation in 
magnitude would cause the following change in $P_{\rm{CW}}$: 

\begin{equation}\label{s45}
\delta P_{\rm{CW}}= \frac{H}{K_Y}\frac{\overline{y}^{(H-1)}}{(1+\overline{y}^{H})^2}\sigma_Y
\end{equation}

where $y=Y/K_Y$ and $\overline{y}$ is its mean value in steady state. The function $\overline{y}^{H-1}/(1+\overline{y}^H)^2$ has a single 
maximum (and no minimum) at $\overline{y}=[(H-1)/(H+1)]^{1/H}\equiv \tilde{y}$, with maximum value 1/4 for $H\gg 1$. Using this property in 
(\ref{s45}) leads to the following upper bound on $\delta P_{\rm{CW}}$: 

\begin{equation}\label{s46}
\delta P_{\rm{CW}}\leq \frac{H}{4K_Y}\sigma_Y
\end{equation}

The standard deviation $\sigma_Y$ has been measured in experiments, and the range of variation has been estimated\cite{bibPark_Nature} as

\begin{equation}\label{s46+}
0.09<\sigma_Y<0.22,
\end{equation}

with the actual value depending on the mean clockwise bias, ${\overline P}_{\rm CW}$. 
The dissociation 
constant $ K_Y \simeq 3\mu M $ and the Hill coefficient $H\simeq 20$ (for pre-adapted motors)
have also been measured in single cell experiments \cite{bibYuan_JMB}. 
Using this estimate in (\ref{s46}), we find $\delta P_{\rm CW} \leq 0.15-0.367$ (assuming $\overline{y}\approx \tilde{y}$).
So there is a maximum fluctuation of $ 18 \% $ for $H=20$ corresponding to a CW bias of $0.5$ and [CheYp] $\sim 3 \mu M$ 
around which the transition becomes ultrasensitive.
Therefore, the fluctuations in [CheYp] appear sufficiently large to affect the switching behavior of the motor.

Next, how large the fluctuations in activity need to be, to produce the observed standard deviation in [CheYp]? The following simple picture 
helps us to obtain an estimate. In {\it E.coli}, active receptors undergo autophosphorylation, and then transfer the phosphoryl groups to 
the proteins CheB and CheY. For simplicity, we ignore phosphorylation of CheB and focus on CheY.  
The dynamics of $Y(t)$ is expressed in (\ref{s41}). Under steady state conditions, 
the variance $\sigma_Y^2=\langle Y^2\rangle-\langle Y\rangle^2$ is given by the expression

\begin{equation}\label{s42}
\sigma_Y^2=\frac{(a_{Y}Y_0A_0)^2}{\lambda_Y(\lambda_Y+\lambda_\xi)}\sigma_a^2
\end{equation}

where, we have introduced the correlation length $\lambda_{\xi}$ for the fluctuations in $\xi_a$, defined through the relation $\langle \delta\xi_a(t)\delta\xi_a(t^{\prime})\rangle=\sigma_a^2 \exp(-\lambda_{\xi}|t-t^{\prime}|)$ in steady state, where $\delta\xi_a(t)=\xi_a(t)-\overline{\xi_a}$ is the fluctuation in activity relative to its steady state value. From (\ref{s42}), the inequality 

\begin{equation}\label{s43}
\sigma_Y\leq\frac{a_{Y}Y_0A_0}{\lambda_Y}\sigma_a
\end{equation}

follows, the equality being satisfied in the limit $\lambda_{\xi}\ll\lambda_Y$ (which we shall assume henceforth). Various estimates 
for the parameters in (\ref{s43}) are available in the literature (see Table\ref{table2}). 
Using (\ref{s46+}), we identify the following ranges for the standard deviation $\sigma_a$ of receptor activity:

\begin{eqnarray}\label{s47}
4.7 \times 10^{-3} & < & \sigma_a <  10^{-2} \;\;\;\;\;\;\;\;\;\;\;\;\; \rm{(i)}\nonumber\\
3 \times 10^{-4} & < & \sigma_a < 7.4 \times 10^{-4} \;\;\; \rm{(ii)}\nonumber\\
5.2 \times 10^{-4} & < & \sigma_a < 10^{-3} \;\;\;\;\;\;\;\;\;\;\; \rm{(iii)}
\end{eqnarray}

It is also of interest to compare the values of the ZOU parameter $\alpha$ across these sets of parameters, which turn out to be 0.547 (i), 
0.076 (ii) and 0.035 (iii) (Table\ref{table2}). Note that the estimated variance in (ii) and (iii) are smaller than (i), which is consistent 
with our expectation of maximum variance near $\alpha=1$, over a large range of values of $\ell$. The magnitude of receptor noise predicted by (\ref{s47}) also agrees with our predictions (see Fig (\ref{fig3}), also Fig (\ref{fig8})). 

\begin{table}[!ht]

\setlength{\tabcolsep}{2 pt} 
\renewcommand{\arraystretch}{1.5}
\caption{\bf{A list of relevant parameter values collected  from the literature.}}
\centering
\begin{tabular}{|p{1.3cm}|p{4.8cm}|p{2.5cm}|p{2.5cm}|p{2.5cm}|}
\hline
\hline
& & \multicolumn{3}{|c|}{Numerical values} \\
\hline
Symbol  & Quantity & Morton-Firth {\it et al.} \cite{bibMorton} & Rao {\it et al.}\cite{bibRao_JCP} & Kollmann {\it et al.}\cite{bibKollmann} \\ 
  
\hline \hline
 $a_Y$ & CheY phosphorylation rate & 3 $\mu M^{-1}s^{-1}$ & 100 $\mu M^{-1}s^{-1}$ & 100 $\mu M^{-1}s^{-1}$ \\
  
 $\lambda_Y$ & CheYp dephosphorylation rate &  14.15 $s^{-1}$ & 30.1 $s^{-1}$ & 30.1 $s^{-1}$ \\ 
 
 $Y_0$ & CheY concentration & 18 $\mu$M  & 17.9 $\mu$M & 9.7 $\mu$M  \\ 
 
 $A_0$ & CheA concentration & 5 $\mu$M & 5 $\mu$M & 5.3 $\mu$M \\
 
 $R_0$ & CheR concentration & 0.235 $\mu$M & 0.3 $\mu$M & 0.16 $\mu$M \\
 
 $B_0$ & CheB concentration & 2.27 $\mu$M & 2 $\mu$M & 0.28 $\mu$M \\ 
 
 $\nu_r$ & Methylation rate & 0.819 $s^{-1}$ & 0.255 $s^{-1}$ & 0.39 $s^{-1}$ \\ 
 
 $\nu_b$ & Demethylation rate & 0.155 $s^{-1}$ & 0.5 $s^{-1}$ & 6.3 $s^{-1}$ \\ 
 
 $\alpha (\ref{alpha}) $ & ZOU parameter & 0.547 & 0.076 & 0.035 \\
\hline \hline
 
\end{tabular}
\begin{flushleft}  
\end{flushleft}
\label{table2}
\end{table}

\subsection{Effect of flagellar motor level adaptation on fluctuations in clockwise bias}
\label{S5_Appendix}
\begin{figure}[!h] 
\begin{center}
\includegraphics[width=8cm, angle = 0]{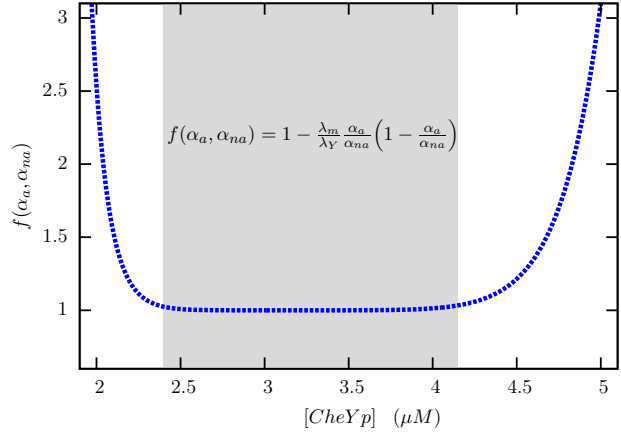}
 \caption{The term inside the square root in (\ref{final}) is plotted against $Y$. Note the flat region, nearly 1.5$\mu M$ in width, 
 where the correction arising from motor-level adaptation is negligible. $\alpha_a$ and $\alpha_{na}$ have been computed from (\ref{s44})
 with $H_a = 10 $ \cite{bibCluzel} and $H_{na} = 20$ \cite{bibYuan_JMB}. We have taken $\lambda_m = 0.0167 s^{-1}$ considering that motor adaptation
 takes place over a minute \cite{bibYuan_Nature} and $\lambda_Y = 30 s^{-1}$ \cite{bibRao_JCP, bibKollmann}.} 
 \label{fig10}
 \end{center}
 \end{figure}

From the experimental data of Yuan and Berg (2013) \cite{bibYuan_JMB}, we propose that the change in clockwise bias in response to a 
time-dependent change in [CheYp] (denoted $\delta Y(t)$ henceforth) be written in the general linear form 
\begin{equation}
\delta P_{\rm CW}(t)=\int_{-\infty}^{t}\chi_m(t-t^{\prime})\delta Y(t^{\prime})dt^{\prime}
\label{linear_response}
\end{equation}
where $\chi_m(\tau)$ (defined for $\tau>0$) is the linear response function for the motor, which we assume to consist of two terms:
\begin{equation}
\chi_m(\tau)=\alpha_{\rm na}\delta(\tau)+\Gamma e^{-\lambda_m\tau}
\label{chim}
\end{equation}
where $\alpha_{\rm na}=\partial_Y P_{\rm CW}^{\rm na}$, and $P_{\rm CW}^{\rm na}$ is the bias of the non-adapted motor, as 
measured by Yuan and Berg (2013), with Hill coefficient $H\approx 20$. The second term represents the slow change in the bias 
originating from motor-level adaptation, with $\lambda_m^{-1}$ being the time scale for the same. Now, for a step-like change 
in $Y$ such that $Y\to Y_0+\Delta Y$ at $t=0$, let us put $\delta Y(t)=\Delta Y\Theta(t)$; using this in (\ref{linear_response}) 
yields the response for the same:
\begin{equation}
\delta P_{\rm CW}^{\rm step}(t)=\Delta Y\int_{0}^{t}\chi_m(\tau)d\tau
\label{step}
\end{equation}
As $t\to\infty$, the system will adapt, and hence $\delta P_{\rm CW}^{\rm step}(t)\to \alpha_{\rm a}\Delta Y$, where 
$\alpha_{\rm a}=\partial_Y P_{\rm CW}^{\rm a}$, and $P_{\rm CW}^{\rm a}$ is the clockwise bias of the {\it adapted} motor. 
From (\ref{chim}), it then follows that $\Gamma=\lambda(\alpha_{\rm a}-\alpha_{\rm na})$. 
Let us now compute the effect on the CW bias of random fluctuations in $Y$, and define the root-mean square change of bias 
by $\delta P_{\rm CW}\equiv \sqrt{\langle \delta P_{\rm CW}(t)\rangle^2}$, with the average to be carried out over different 
realizations of the fluctuations in [CheYp]. From (\ref{linear_response}), it turns out that 
\begin{equation}
\langle \delta P_{\rm CW}(t)\rangle^2=\int_{-\infty}^{t}dt_1\int_{-\infty}^{t}dt_2 \chi_m(t-t_1)\chi_m(t-t_2)\langle \delta Y(t_1)\delta Y(t_2)\rangle,
\label{integral}
\end{equation}
where, in steady state, we expect $\langle \delta Y(t_1)\delta Y(t_2)\rangle=\sigma_Y^2 e^{-\lambda_Y(t_1-t_2)}$ for $t_1>t_2$, $\sigma^2_Y$ is the steady state variance of $Y$ as given by (\ref{s42}) and $\lambda_Y$ is the dephosphorylation rate of [CheYp]. Using the expression for 
$\chi_m(t)$ from (\ref{chim}) and completing the calculations yields the final result
\begin{equation}
\delta P_{\rm CW}=\sigma_Y\alpha_{\rm na}\sqrt{ \left[1-\frac{\lambda_m}{\lambda_Y}\frac{\alpha_{\rm a}}{\alpha_{\rm na}}\left(1-\frac{\alpha_{\rm a}}{\alpha_{\rm na}}\right)\right]}
\label{final}
\end{equation}
The term outside the square root is precisely the result in (\ref{s45}) in \ref{S3_Appendix}. Explicit evaluation of the term 
in brackets (see Fig.\ref{fig10}) shows, however, that the contribution of motor-level adaptation to the fluctuations in bias is 
negligible in the sensitive part of the response curve.  

\section*{Acknowledgments}
We thank P.G. Senapathy Centre for Computing Resources, IIT Madras, for providing us with computing facilities in the Virgo cluster.


\end{document}